\documentclass[english]{IEEEtran}
\usepackage[T1]{fontenc}
\usepackage[utf8x]{inputenc}
\usepackage{array}
\usepackage{textcomp}
\usepackage{graphicx}

\makeatletter

\DeclareFontEncoding{LGR}{}{}
\DeclareTextSymbol{\~}{LGR}{126}
\providecommand{\tabularnewline}{\\}

\usepackage[utf8x]{inputenc}
\usepackage[T1]{fontenc}
\usepackage{textcomp}
\usepackage{cite}

\makeatother

\usepackage{babel}
\begin{document}

\title{On the Security of the Automatic Dependent Surveillance-Broadcast
Protocol}

\author{Martin Strohmeier$^{*}$, Vincent Lenders$^{+}$, Ivan Martinovic$^{*}$\\
$^{*}$University of Oxford, United Kingdom\\
$^{+}$armasuisse, Switzerland}
\maketitle
\begin{abstract}
Automatic dependent surveillance-broadcast (ADS-B) is the communications
protocol currently being rolled out as part of next generation air
transportation systems. As the heart of modern air traffic control,
it will play an essential role in the protection of two billion passengers
per year, besides being crucial to many other interest groups in aviation.
The inherent lack of security measures in the ADS-B protocol has long
been a topic in both the aviation circles and in the academic community.
Due to recently published proof-of-concept attacks, the topic is becoming
ever more pressing, especially with the deadline for mandatory implementation
in most airspaces fast approaching. 

This survey first summarizes the attacks and problems that have been
reported in relation to ADS-B security. Thereafter, it surveys both
the theoretical and practical efforts which have been previously conducted
concerning these issues, including possible countermeasures. In addition,
the survey seeks to go beyond the current state of the art and gives
a detailed assessment of security measures which have been developed
more generally for related wireless networks such as sensor networks
and vehicular ad hoc networks, including a taxonomy of all considered
approaches.\end{abstract}

\section{Introduction}

\thispagestyle{empty}The world of air traffic control (ATC) is moving
from uncooperative and independent (primary surveillance radar, PSR)
to cooperative and dependent air traffic surveillance (secondary surveillance
radar, SSR). This paradigm shift holds the promise of reducing the
total cost of deployment and improving the detection accuracy of aircraft.
However, it is well known in the aviation community that the ATC system,
which is currently being rolled out, called \emph{automatic dependent
surveillance-broadcast} (ADS-B), has not been developed with security
in mind and is susceptible to a number of different radio frequency
(RF) attacks. The problem has recently been widely reported in the
press \cite{NewScientist,CNN,Forbes,Wired,NPR} and at hacker conventions
\cite{Costin,bradhaines2012,defconkunkel}. Academic researchers,
too, proved the ease of compromising the security of ADS-B with current
off-the-shelf hard- and software \cite{schafer2013experimental}.
This broad news exposure led the International Civil Aviation Organization
(ICAO) to put the security of civil aviation on their agenda of the
12th air navigation conference, identifying ``cyber security as a
high-level impediment to implementation that should be considered
as part of the roadmap development process'' \cite{ICAO2012} and
creating a task force to help with the future coordination of the
efforts of involved stakeholders. 

This shows that there is a widespread concern about the topic, created
by the fact that ADS-B will be mandatory for all new aircraft in the
European airspace by 2015%
\footnote{Older aircraft need to be retrofit by 2017. The FAA mandates ADS-B
in the US airspace by 2020.%
} and has already been embraced by many airlines worldwide. Reports
from manufacturers and regulation bodies show that around 70-80 percent
of commercial aircraft worldwide have been equipped with ADS-B transponders
as of 2013 \cite{adsbimplementation,airbusstatus}. Countries such
as Australia have already deployed full continental coverage, with
ADS-B sensors being the single means of ATC in low population parts
of the country \cite{cascade}.

This paper gives an overview of the research that has been conducted
regarding the security of ADS-B and describes the potential vulnerabilities
identified by the community. Since much relevant security research
has been conducted in related fields such as wireless sensor networks
or general ad hoc networks, we analyze proposed countermeasures from
other areas that could be adapted for use in ADS-B or whether they
are not applicable for reasons inherent to the system. Furthermore,
the present survey provides a threat catalogue, analysis and vulnerability
categorization of the mainly used data link \textit{Mode S}. We focus
primarily on the security of ADS-B and not on other air traffic control
(sub-)\,systems, such as GPS. Among other questions, we seek to answer
why existing ideas for securing (wireless) networks such as traditional
cryptography cannot simply be transferred and used in the protection
of systems such as ADS-B.

While there were a number of reasons behind the switch to a modern
air traffic management system, cost has consistently been mentioned
as one of the most important ones throughout the process; existing
radar infrastructures are simply much more expensive to deploy and
maintain \cite{Smith2006}. ADS-B, on the other hand, provides significant
operational enhancements for both airlines and air traffic managers.
The increased accuracy and precision improves safety and decreases
the likelihood for incidents by a large margin, unless the system's
weak security is exploited by malicious in- and outsiders.

The remainder of the survey is organized as follows: Section \ref{sec:Problem-Statement}
gives a detailed overview over the security problems related to ADS-B
and the requirements of the system environment. Section \ref{sec:Outlining-Solutions}
outlines solutions proposed in previous works and looks at the sister
protocol ADS-C and military versions of ADS-B. Section \ref{sec:Secure-Broadcast-Authentication}
surveys secure broadcast authentication methods to address the problem
while Section \ref{sec:Secure-Location-Verification} reviews means
to establish secure location verification with ADS-B. Section \ref{sec:Summary}
summarizes and Section \ref{sec:Conclusion} concludes the survey.

\section{Problem Statement \label{sec:Problem-Statement}}

This section defines the problems related to security in ADS-B more
thoroughly. First, we give a short overview over the currently used
ADS-B protocol and its existing vulnerabilities. Building on this,
a model of the ADS-B environment is outlined and the required attributes
of possible solutions are identified.\\

\subsection{ADS-B Protocol Overview}

The American Federal Aviation Administration (FAA) as well as its
European pendant EUROCONTROL named ADS-B as the satellite-based successor
of radar. At its introduction, ADS-B was a completely new paradigm
for air-traffic control. Every participant retrieves their own position
and velocity by using an onboard GPS receiver. The position is then
periodically broadcasted in a message (typically twice per second)
by the transmitting subsystem called ADS-B Out.\emph{ }The messages
are then received and processed by ATC stations on the ground as well
as by nearby aircraft, if equipped with the receiving subsystem ADS-B
In.\emph{ }Messages can integrate further fields such as ID, intent,
urgency code, and uncertainty level.\emph{ }

\begin{figure}
\includegraphics[scale=0.75]{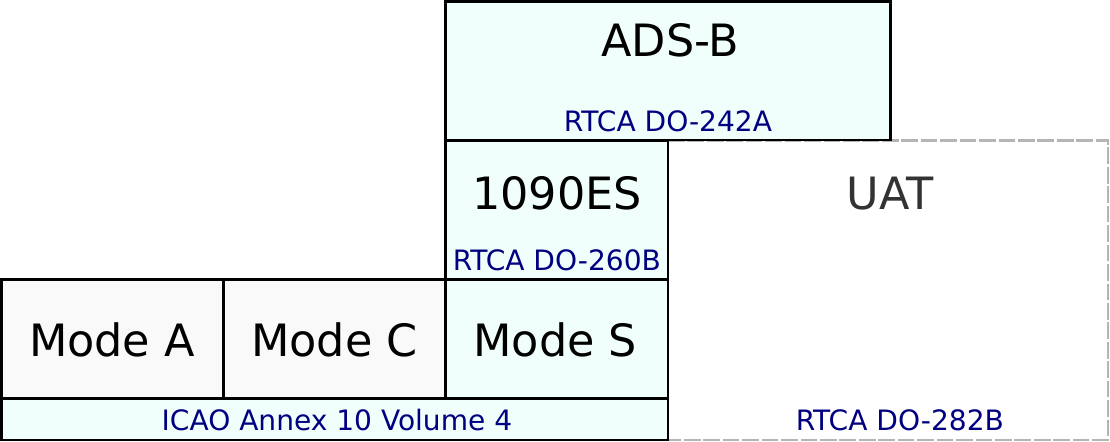}\caption{ADS-B hierarchy \cite{schafer2013experimental}.\label{fig:ADS-B-hierarchy.}
The 1090 MHz Extended Squitter is based on the traditional Mode S
system and provides the data link for ADS-B in commercial aviation.
UAT is a new development but currently only mandated for general aviation. }
\end{figure}

Two competing ADS-B data link standards have been proposed: Universal
Access Transceiver (UAT) and 1090 MHz Extended Squitter (1090ES).
UAT has been created specifically for the use with aviation services
such as ADS-B, utilizing the 978MHz frequency with a bandwidth of
1Mbps. Since UAT requires fitting new hardware, as opposed to 1090ES,
it is currently only used for general aviation\emph{ }in EUROCONTROL
and FAA-mandated airspaces. Commercial aircraft, on the other hand,
employ SSR Mode S with Extended Squitter, a combination of ADS-B and
traditional Mode S known as 1090ES (see Fig.\,\ref{fig:ADS-B-hierarchy.}).\emph{
}In other words, the ADS-B function can be integrated into traditional
Mode S transponders. From here on, we focus on the commercially used
1090ES data link. The complete overview over the ADS-B protocol can
be found in the specification documents \cite{DO2422,DO260B,DO282B2}
while various other works give succinct, higher level descriptions
of the protocol (e.g.\,\cite{Costin,schafer2013experimental,McCallie2011}).

\subsubsection*{The 1090ES Data Link}

\begin{figure}
\includegraphics[scale=0.5]{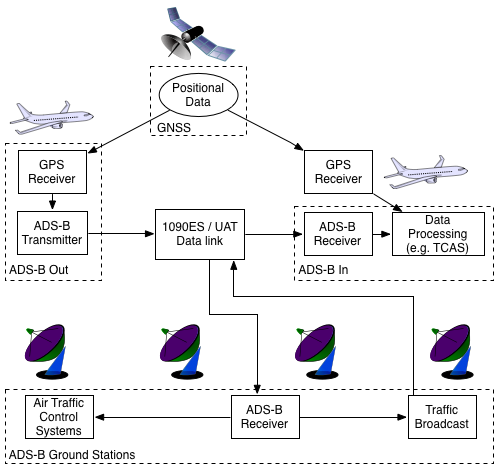}

\caption{Overview of the ADS-B system architecture. Aircraft receive positional
data that is transmitted via the ADS-B Out subsystem over the 1090ES
or the UAT data link. It is then received and processed by ground
stations and by other aircraft via the ADS-B In subsystem. \label{fig:Overview-ADS-B}}
\end{figure}

\begin{figure}
\includegraphics[scale=0.65]{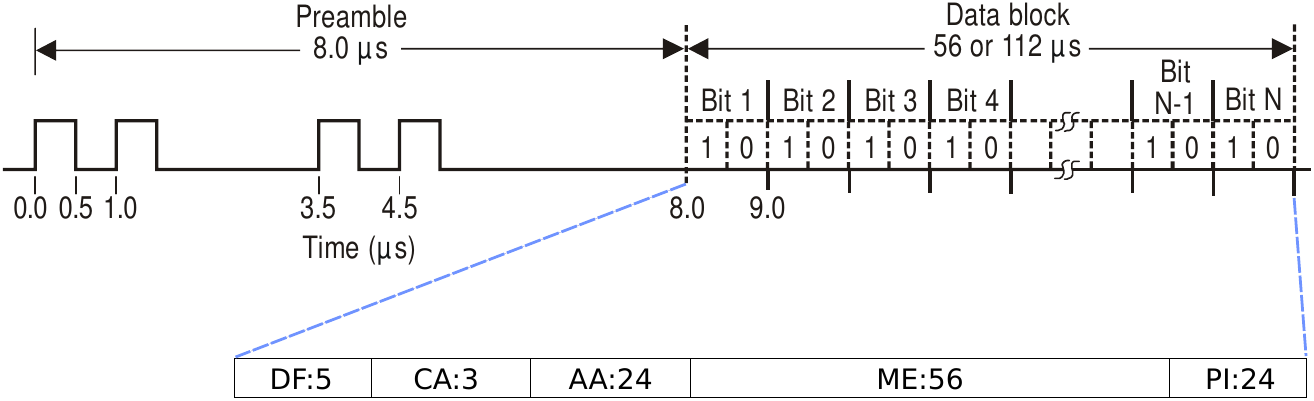}

\caption{1090 ES Data Link \cite{schafer2013experimental}\label{fig:1090-ES-Data}}
\end{figure}

As the name suggests, the 1090ES data link predominantly uses the
1090MHz frequency for communication sent out by aircraft, to both
other aircraft and ground stations\emph{ }(Mode S also uses ground
to air communication at 1030MHz for interrogations and information
services). Figure \ref{fig:1090-ES-Data} provides a graphical view
of a 1090ES transmission, which starts off with a preamble of two
synchronization pulses. The data block is then transmitted by utilizing
pulse position modulation (PPM). With every time slot being 1\textmu{}s
long, a bit is indicated by either sending a 0.5\textmu{}s pulse in
the first half of the slot (1-bit) or in the second half (0-bit).
It is important to note that PPM is very sensitive to reflected signals
and multipath dispersion, a fact that can play a major role in security
and protocol considerations.%
\footnote{See \cite{barry} for more information on PPM and multipath.%
}

There are two different possible message lengths specified in Mode
S, 56 bit and 112 bit \cite{DO2422}, whereas ADS-B solely uses the
longer format. The downlink format field DF (alternatively UF for
uplink messages) assigns the type of the message.\emph{ }1090ES uses
a multipurpose format as shown in Fig.\,\ref{fig:1090-ES-Data}.
When set to 17, it indicates that the message is an extended squitter,
enabling the transmission of 56 arbitrary bits in the ME field.\emph{
}The CA field indicates information about the capabilities of the
employed transponder, while the\emph{ }24 bit AA field carries the
unique ICAO aircraft address which enables aircraft identification.\emph{
}Finally, the PI-field provides a 24 bit CRC to detect and correct
possible transmission errors.\emph{ }It is possible for recipients
to correct up to 5 bit errors in 1090ES messages using a fixed generator
polynomial of degree 24.

This quick overview shows that only the 56 bit ME field can be used
to transmit arbitrary data, i.e. utilized for a secure ADS-B solution.
However, not only is it very limited in size but it is also typically
occupied by positional and other data. Thus, the format as currently
in practical use is intuitively very limiting to most types of security
solutions as we will explain in more detail in this survey.

\subsection*{Relation to Legacy Systems}

Traditionally, aircraft localization has been relying on radar systems
which had been developed for military applications, namely identification,
friend or foe (IFF) systems. There are two different concepts in conventional
radars: primary surveillance radars and secondary surveillance radars
\cite{Skolnik07}. PSRs are \textit{independent}; they work without
cooperation from the aircraft by transmitting high-frequency signals,
which the target object reflects. The echo identifies range, angular
direction, velocity, size and shape of the object. SSR, on the other
hand, uses interrogations from ground stations which are responded
to by transponders in aircraft. The reply includes information such
as the precise aircraft altitude, identification codes or information
about technical issues. In contrast to PSR, this approach is also
much more accurate in terms of localization and identification. As
all surveillance data such as position and status are derived directly
by the aircraft, SSR is \textit{dependent}. Furthermore, cooperation
by the aircraft is a requirement.

Before ADS-B, all SSR systems in ATC have been interrogation-based.
So called \textit{modes} are being used to query the identification
and altitude of an aircraft. There are three modes (A, C and S) currently
in use in civil aviation, Table \ref{tab:Comparison-of-different}
compares their characteristics with ADS-B. The latter embodies a paradigm
shift in ATC as air traffic surveillance is now cooperative and dependent,
i.e.~every aircraft collects their own data such as position and
velocity by using onboard measurement devices. ADS-B based surveillance
infrastructure is also much more cost-effective compared to conventional
PSR, which is a much more complex technology that also suffers from
wear and tear due to rotating parts. ICAO specifies the technological
cost of using primary radar to monitor an en-route airspace (200NM
radius) at \$10-14 million, while Mode S surveillance is priced at
\$6 million and ADS-B is significantly cheaper at \$380,000 \cite{ICAO2007}.
\begin{table}
\begin{tabular}{|>{\centering}p{1.1cm}|>{\centering}p{1.5cm}|>{\centering}p{1.2cm}|>{\centering}p{1.6cm}|>{\centering}p{1.3cm}|}
\hline 
 & Message Length  & Frequencies  & Operational Mode  & Use Cases\tabularnewline
\hline 
\hline 
Mode A & 12 bit & 1030 / 1090\,MHz & Independent / Non-selective interrogation  & Identification\tabularnewline
\hline 
Mode C & 12 bit & 1030 / 1090\,MHz & Independent / Non-selective interrogation  & Pressure Altitude Extraction \tabularnewline
\hline 
Mode S & 56 / 112 bit & 1030 / 1090\,MHz & Independent / Selective interrogation  & Multiple\tabularnewline
\hline 
ADS-B / 1090\,ES & 112 bit & 1090\,MHz & Dependent / Automatic  & Multiple\tabularnewline
\hline 
\end{tabular}

\caption{Comparison of civil aviation transponder modes \cite{Strohmeier14}.
\label{tab:Comparison-of-different}}
\end{table}

\subsection{ADS-B Vulnerabilities}

In this section, we discuss the ADS-B vulnerabilities inherently stemming
from the broadcast nature of RF communication when used without additional
security measures. Contrary to wired networks, there are no practical
obstacles for an attacker trying to access a wireless network such
as buildings or security guards\emph{,} making access control mechanisms
very challenging. In \cite{McCallie2011}, McCallie et al.~defined
a taxonomy for various possible attacks, even though their difficulty
estimations have become somewhat dated with the widespread availability
of cheap software-defined radios as recently illustrated in \cite{schafer2013experimental}.
Despite this, the described attacks are ordered in increasing order
of difficulty here, providing a comprehensive attacker model in the
context of ADS-B vulnerabilities.\\

\paragraph*{Eavesdropping}

The most straightforward form among the many security vulnerabilities
present in ADS-B is the act of listening in to the unsecured broadcast
transmissions. This passive attack is called \emph{Aircraft Reconnaissance}
in \cite{McCallie2011}. As ADS-B is using unsecured messages over
an inherently broadcast medium, the possibility to eavesdrop is not
surprising and has been mentioned since the early stages of development.
Many non-adversarial services use this obvious privacy concern, e.g.,
to visualize air-traffic on the Internet,%
\footnote{Prominent examples are flightradar24.com and radarvirtuel.com among
many others.%
} yet eavesdropping also forms the basis for a number of more sophisticated
active attacks. Furthermore, eavesdropping is not only difficult to
prevent without applying full encryption but it is also practically
impossible to detect. A small number of countries (such as the United
Kingdom) have long-standing, very general laws against listening in
on unencrypted broadcast traffic which is not intended for the recipient%
\footnote{Section 48 of the Wireless Telegraphy Act of 2006 states that (1)
``A person commits an offence if, otherwise than under the authority
of a designated person— (a) he uses wireless telegraphy apparatus
with intent to obtain information as to the contents, sender or addressee
of a message (whether sent by means of wireless telegraphy or not)
of which neither he nor a person on whose behalf he is acting is an
intended recipient, or (b) he discloses information as to the contents,
sender or addressee of such a message.''%
} even though the technical realities render such legal approaches
all but obsolete.\\
\begin{table*}
\begin{centering}
\begin{tabular}{|>{\centering}p{3.2cm}|>{\centering}p{1.7cm}|>{\centering}p{1.7cm}|>{\centering}p{1.7cm}|>{\centering}p{1.7cm}|>{\centering}p{1.7cm}|}
\hline 
Attacks & Layer  & Method  & Severity  & Complexity & References\tabularnewline
\hline 
\hline 
Aircraft Reconnaissance\\
 & PHY+APP & Eavesdropping & Low & Lowest & \cite{Costin,schafer2013experimental,McCallie2011}\tabularnewline
\hline 
Ground Station Flood Denial & PHY & Signal Jamming & Medium & Lower & \cite{Costin,schafer2013experimental,McCallie2011,Wilhelm2011a,Purton2010}\tabularnewline
\hline 
Aircraft Flood Denial & PHY & Signal Jamming & Medium & Low-Medium & \cite{Costin,schafer2013experimental,McCallie2011,Wilhelm2011a,Purton2010}\tabularnewline
\hline 
Ground Station Target Ghost Injection / Flooding & APP & Message Injection & High & Low & \cite{Costin,schafer2013experimental,McCallie2011}\tabularnewline
\hline 
Aircraft Target Ghost Injection / Flooding & APP & Message Injection & Medium & Low-Medium & \cite{Costin,schafer2013experimental,McCallie2011}\tabularnewline
\hline 
Virtual Aircraft Hijacking & PHY+APP & Message Modification & High & Medium & \cite{schafer2013experimental,Popper2011,Wilhelm2012}\tabularnewline
\hline 
Virtual Trajectory Modification & PHY+APP & Message Modification & High & Medium & \cite{schafer2013experimental,Popper2011,Wilhelm2012}\tabularnewline
\hline 
Aircraft Disappearance & PHY & Message Deletion & High & Low & \cite{schafer2013experimental,McCallie2011}\tabularnewline
\hline 
Aircraft Spoofing & PHY+APP & Message Modification & High & Low & \cite{Costin,schafer2013experimental,Purton2010}\tabularnewline
\hline 
\end{tabular}
\par\end{centering}

\caption{Overview of vulnerabilities in the ADS-B protocol. The Table summarizes
the attacks including severity and complexity. \label{tab:Comparison-of-vulnerabilities}}
\end{table*}

\paragraph*{Jamming}

Almost equally simple is the jamming attack, where a single node (both
ground stations or aircraft) or an area with multiple participants
is effectively disabled from sending/receiving messages by an adversary
sending with sufficiently high power on the 1090MHz frequency of Mode
S. It has generally also been proven feasible to do reactive jamming
in real time, targeting only packets which are already in the air
as assessed in \cite{Wilhelm2011a}. While jamming is a problem common
to all wireless communication, the impact is severe in aviation due
to the system's inherent wide open spaces which are impossible to
control as well as the importance and criticality of the transmitted
data. Besides SSR communications systems such as ADS-B, primary radar
may also be a target of jamming attacks which is similar in many ways,
most importantly the fact that one always jams the receiver (in our
case ATC systems), not the transmitter.%
\footnote{Of course, PSR does have both a transmitter and a receiver, while
ADS-B ground stations are dependent on the tracked targets' transmissions.%
} A thorough introduction to radar and communication jamming can be
found in \cite{adamy2001ew}.

Due to rotating antennas and a higher transmission power, typical
PSRs are more difficult to jam than ADS-B receivers, especially for
non-military grade attackers. However, as there are usually many distributed
ADS-B receivers for ATC purposes, it still takes considerable effort
to completely blackout a given area. Notwithstanding this, a targeted
attack would create major denial-of-service problems at any airport.
Jamming moving aircraft is also possible, however considered more
difficult. In summary, jamming is integral for the following attacks\cite{McCallie2011}:

\vspace{+0.1in}
\begin{itemize}
\item Ground Station Flood Denial
\item Aircraft Flood Denial\\

\end{itemize}

\paragraph*{Message Injection}

On the next higher level of difficulty, it is also possible to inject
non-legitimate messages\emph{ }into the air-traffic communication
system. Since no authentication measures are implemented at the data
link layer, there is no hurdle at all for an attacker to build a transmitter
that is able to produce correctly modulated and formatted ADS-B messages.
See \cite{schafer2013experimental} for more details on how to conduct
an attack with limited knowledge and very cheap and simple technological
means which have been easily and widely available for some time. As
another direct consequence of missing authentication schemes, a node
can deny having broadcasted any (false) data and/or claim having received
conflicting data, making any kind of liability impossible. Concrete
attack instances that use message injection include \cite{McCallie2011}:

\vspace{+0.1in}
\begin{itemize}
\item Ground Station Target Ghost Injection/Flooding
\item Aircraft Target Ghost Injection/Flooding\\

\end{itemize}

\paragraph*{Message Deletion}

Legitimate messages can be physically ``deleted'' from the wireless
medium by utilizing\emph{ }destructive or constructive interference.\emph{
}Destructive interference means transmitting the inverse of the signal
broadcast by a legitimate sender.\emph{ }Due to superposition, the
resulting signal should be erased or at least highly attenuated but
in practice this approach has very precise and complex timing requirements,
making it extremely challenging.\emph{ }

Constructive interference on the other hand does not require synchronization
but simply causes a large enough number of bit errors.\emph{ }Since
Mode S extended squitters' CRC can correct a maximum of 5 bit errors
per message, if a message exceeds this threshold, the receiver will
drop it as corrupted. While effectively destroyed, the receiver might
at least be able to verify that a message has been sent, depending
on the implementation and the circumstances. In any case, it is more
subtle than complete jamming of the 1090MHz frequency. Besides providing
an easy means of message modification in conjunction with message
injection, message deletion is key to the following attack:

\vspace{+0.1in}
\begin{itemize}
\item Aircraft Disappearance\\

\end{itemize}

\paragraph*{Message Modification}

Modifying messages on the physical layer during transmission is typically
done via two different approaches,\emph{ }overshadowing and bit-flipping.\emph{
}Overshadowing means that the attacker sends a high-powered signal
to replace part or all of the target message.\emph{ }Bit-flipping
on the other hand has the attacker superimposing the signal converting
any number of bits from 1 to 0 (or the other way around). In both
cases arbitrary data can be injected without the knowledge of any
of the participants. This effect can also be achieved by combining
message deletion and injection, but physical layer message modification
can in some cases be regarded as even more sinister than the injection
of a completely new message, since the manipulated message was originally
legitimate. The feasibility of such message manipulation has recently
been shown in \cite{Popper2011} and \cite{Wilhelm2012}. Concrete
attack examples are \cite{schafer2013experimental}:

\vspace{+0.1in}
\begin{itemize}
\item Virtual Aircraft Hijacking
\item Virtual Trajectory Modification\\

\end{itemize}

\subsection{Identification of System Requirements \label{sub:Requirements}}

There are a number of demands on a security approach for ADS-B, stemming
from the way it is needed to work in practical real-world aviation
settings and the characteristics of the broadcast approach. We first
codify the model and then follow up with an analysis of what we need
to have as security primitives in air traffic control systems.\\

\subsubsection*{Network Properties}
\begin{itemize}
\item The assumed network model consists purely of unidirectional broadcasts.
Although there is a growing body of research on Aeronautical Ad hoc
Networks (AANETs) that provide multi-hop communication \cite{Li2010},
the present real-world implementation is based on \textbf{single-hop
unidirectional broadcast links}. Aircraft broadcast their position,
velocity and direction in plain-text periodically\emph{ }every few
hundred milliseconds, a concept called beaconing. Thus, in the following
we concentrate mainly on the so-called beacon-based security.
\item We \textbf{do not consider any type of energy constraints} in association
with ADS-B devices. 
\item Furthermore, there are \textbf{no significant computational constraints}
with neither ground stations nor ADS-B units employed in aircraft.
\item \textbf{Reliability has not been a major concern} yet, as some lost
packets do not normally cause a problem. Indeed, the protocol does
have no means to prevent collisions, the sender will not retransmit
any packets and no guarantees are given. Loss is dealt with on higher
layers. This is also reflected in \cite{schafer2013experimental},
which shows that the packet error rate tends to hover around a mean
of 33 \%, independent of the channel. This means that there is some
substantial packet loss on the physical layer that is likely to increase
when the channel utilization rises over the next decades due to the
mandatory ADS-B roll-out. This will be reinforced by the ever-increasing
flight traffic,%
\footnote{Growth rate forecasts from market analysts suggest a 5.1\% increase
per year between 2010 and 2030. Cargo traffic is expected to grow
even more at 5.6\% per year in the same time frame \cite{Boeing}. %
} especially in high-density airspaces.
\item \textbf{The network is ad hoc and highly mobile.} Many nodes are moving
at a velocity of up to 1,000 km/h or more. It is therefore extremely
dynamic and communication between two nodes may last only a few seconds.
Aircraft trajectories are principally not physically restricted, although
there are often common routes and also some air spaces that are restricted
due to policies.
\item \textbf{The network is long range}, typically ADS-B is considered
to be feasible at distances of 100 NM and more.%
\footnote{Aviation typically uses the nautical mile, also abbreviated as nmi,
as a distance measure which is 1,852m.%
}
\item In contrast to other, independently deployed wireless (sensor) networks
the \textbf{undetected physical capture of legitimate nodes is not
the most important concern.} Legally having access to a legitimate
ADS-B node, however, is not considered very difficult, at least when
taking general aviation into account.\\

\end{itemize}

\subsubsection*{Security Attributes}

Perrig and Tygar \cite{Perrig2003} identify two large themes around
which secure broadcast revolves: First, make sure that receivers know
that any received information comes from the appropriate sender, and
second that senders can freely limit the recipients of broadcasted
information. Confidentiality, e.g. the protection of ADS-B position
messages to prevent attacks in air or loss of trade secrets, has not
been considered in the development of the system as laid out in a
formal security requirements engineering for ADS-B in \cite{haley2008security}.

In light of the recently exploited inherent vulnerabilities of the
ADS-B system with cheap off-the-shelf hardware \cite{Costin,schafer2013experimental},
this is quickly becoming the most important topic in ADS-B research.
Based on the model defined above, any security scheme for ADS-B would
have to satisfy the following properties:
\begin{itemize}
\item \emph{Data integrity}: Ensures that the data is the same as has been
provided by the sender and has not been modified by any third party.
\item \emph{Source integrity: }Ensures that a message originates from the
participant that claims to have sent it. 
\item \emph{Data origin authentication}: Ensures that a message originates
from the location claimed in a message.
\item \emph{Low impact on current operations:} A scheme should be compatible
with the current ADS-B installations and not overly affect both hard-
and software standards.
\item Sufficiently\emph{ quick and correct detection} of incidents.
\item Needs to be \emph{secure against DoS-attacks} against computing power.
\item Any approach \emph{needs to be easily scalable}. This is in respect
to both a locally rising aircraft density and globally increasing
aircraft traffic. The strain on the heavily used 1030 MHz channel
should not measurably increase (e.g. due to an increased number and/or
larger packets).
\item \emph{Robustness to packet loss.} A jammed wireless channel should
decrease neither security nor reliability of the scheme.
\item Achieving \emph{non-repudiation} is seen as nice to have but not very
high on the priority list for immediate air traffic security and is
more of a legal topic.\\

\end{itemize}

\section{Outlining Solutions \label{sec:Outlining-Solutions}}

Substantial work has already been done on ADS-B security over the
last decade and several approaches have been proposed in the literature
to enhance ADS-B security in particular. Furthermore, a large amount
of research has been done in related fields such as vehicular ad hoc
networks (VANETs) and wireless sensor networks where broadcast authentication
and security also play an important role. While some ideas may not
be directly useful to the aforementioned requirements of ADS-B, it
might be possible to adapt them. We list the major obstacles for their
application, along with the advantages and disadvantages of each approach.

The remainder of this section first presents an overview of works
in the community concerned with the security of ADS-B as a whole.
We then consider ADS-B's sister protocols ADS-C and military versions
of ADS-B in our context before analyzing all collected ideas in detail
in the following sections.

\begin{figure}
\includegraphics[bb=8bp 0bp 719bp 393bp,clip,scale=0.36]{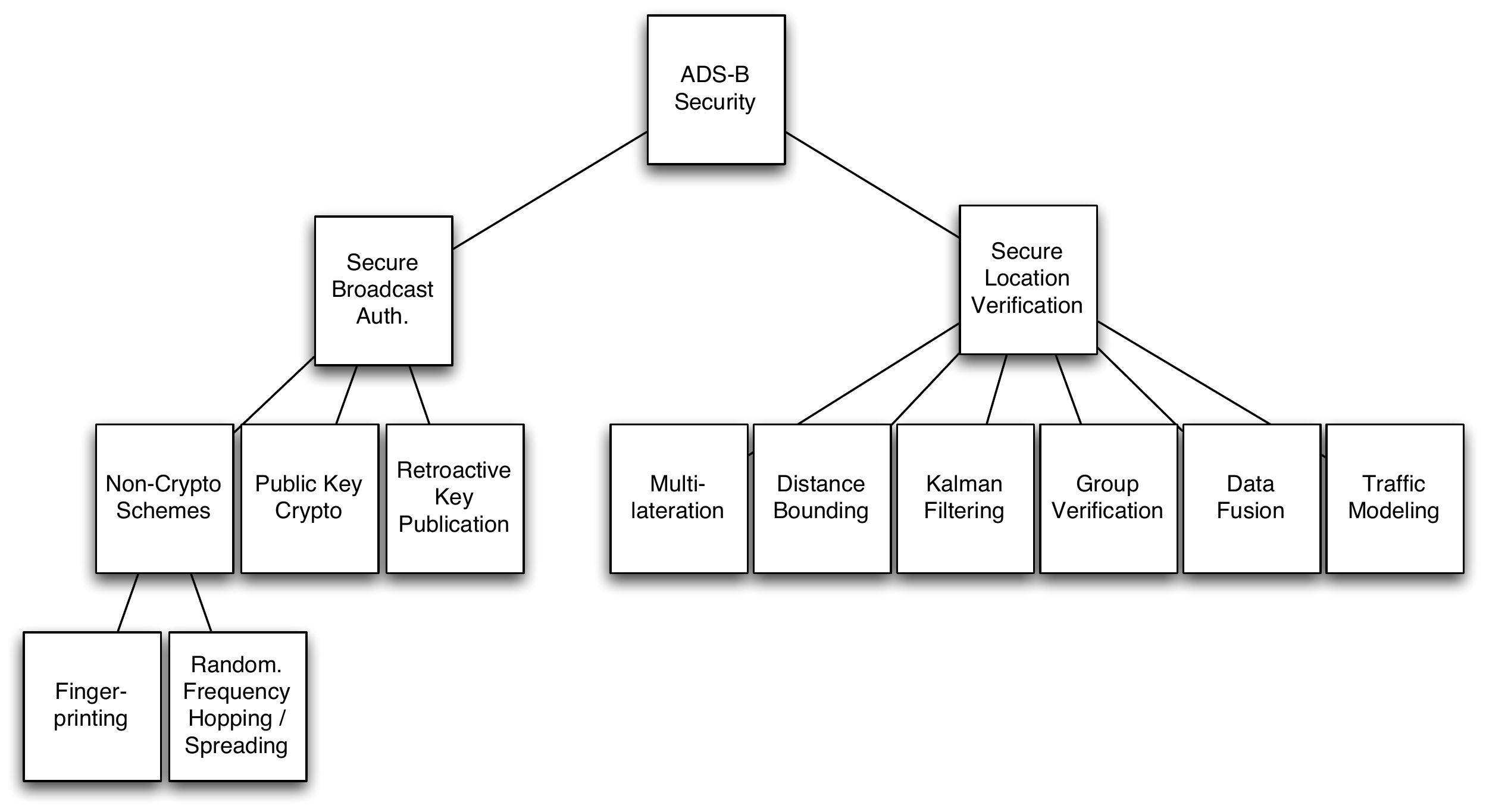}

\caption{Taxonomy of ADS-B Security \label{fig:Systematization-of-ADS-B}}
\end{figure}

As shown in the taxonomy in Fig.\,\ref{fig:Systematization-of-ADS-B},
we identified two distinct approaches to securing ADS-B: \textbf{Secure
Broadcast Authentication} and \textbf{Secure Location Verification}.
Consequently, Section \ref{sec:Secure-Broadcast-Authentication} examines
the various schemes that apply asymmetric properties (cryptographic
and non-cryptographic) to directly authenticate broadcast communication
while Section \ref{sec:Secure-Location-Verification} reviews several
different methods that seek to verify the authenticity of location
claims made by aircraft and other ADS-B participants.

\subsection{Previous Works on ADS-B Security}

Securing ADS-B communication was not a very high priority when it
was specified to be the new standard in civilian secondary surveillance.
Neither the official standards of the Radio Technical Commission for
Aeronautics (RTCA) \cite{DO2422,DO260B,DO282B2} nor other requirements
documents \cite{Martone2001a,Rekkas2008} mention security in this
context. However, security problems in ADS-B have been well-known
for a long time, mostly because they are relatively obvious to the
interested researcher.%
\footnote{Especially on the Internet broad warnings have been floating around
as early as 1999, e.g. http://www.airsport-corp.com/adsb2.htm, http://www.dicksmithflyer.com.au/cat\_index\_36.php%
} For instance, weak ADS-B security has recently got very broad reporting
in the mainstream press%
\footnote{E.g. http://www.forbes.com/sites/andygreenberg/2012/07/25/next-gen-air-traffic-control-vulnerable-to-hackers-spoofing-planes-out-of-thin-air/%
} due to two talks at the DEFCON and Black Hat security conferences.%
\footnote{http://www.blackhat.com/usa/bh-us-12-briefings.html\#Costin%
} 

Sampigethaya and Poovendran \cite{Sampigethaya2011,Sampigethaya}
first analyse the security and privacy of ADS-B and other related
and unrelated communication systems which are part of the so-called
``e-enabled aircraft''. McCallie et al.\,\cite{McCallie2011} provide
a current security analysis, focused on the nature of possible attacks
and their difficulty. They give a systematic high-level overview and
propose general recommendations for addressing ADS-B's problems. Costin
and Francillon \cite{Costin} as well as Schaefer et al.\,\cite{schafer2013experimental}
analyse ADS-B security, too, focusing on the ease of exploiting ADS-B
with current hard- and software and offering some possible countermeasures. 

Kovell et al.\,\cite{Kovell2012} mention data fusion with other
systems, multilateration and various cryptographic schemes as state
of the art research on ADS-B security. They further conduct a more
thorough analysis on Kalman filtering and group validation concepts
and proposed mitigation methods. Nuseibeh et al.~conduct an example
of a formal security requirements analysis of ADS-B \cite{Nuseibeh2009},
proposing multilateration to deal with possible attack scenarios.
Burbank et al.\,\cite{Burbank2005} present general concepts for
communications networking to meet the requirements of future airspace
systems, i.e. a vision of a mobile ad hoc and wireless networking
concept for use in both the terminal area and in the en-route airspace.
Li and Kamal \cite{Li2011} analysed the security of the whole FAA's
Next Generation Air Transportation System (NextGen) of which ADS-B
is a core component.%
\footnote{Although certainly not the only safety-critical wireless module. See
e.g.\,\cite{Ochieng2003} for concerns about the impact of GPS integrity
on aviation safety. %
} They develop a high-level defense-in-depth framework for analyzing
NextGen and mention general secure communication approaches such as
encryption, authentication and spread spectrum as part of a possible
ADS-B security layer that would need to be examined more deeply.

The number of works relating to the security of ADS-B has been increasing
steadily as the final mandate for its use in US and other Western
airspaces is drawing closer and the problem is becoming more urgent.
While the existing works pertain the need for security and offer insights
into different aspects, especially multilateration and data fusion,
the present work seeks to widen the understanding of the communications
community and address the problem on a much broader and more comprehensive
scale. We look to compile all previously considered aspects, seek
to include some applicable, overlooked ideas from other areas of network
security, and weigh the advantages and disadvantages of the examined
approaches against each other.

\subsection{ADS-C}

A theoretical, already available, way to deal with suspicious ADS-B
participants would be to ask them to switch to the connection-oriented
ADS-C (ADS-Contract, also known as ADS-Addressed). However, ADS-C
also has a number of very severe inherent shortcomings, some of which
are described by ICAO: \cite{Siu2011} 
\begin{itemize}
\item Additional avionics systems (for data communications) such as the
Future Air Navigation System (FANS) 1/A or the Aeronautical Telecommunications
Network (ATN) are needed.
\item Its performance may be limited by the communications medium.
\item Currently, ADS-C data is carried by a data link service provider,
so a cost may be incurred for each transmission.
\item Unlike the ADS-B IN system, ADS-C messages are not directly available
to other aircraft.
\end{itemize}
Overall, the current implementation of ADS-C over the outdated ACARS
network%
\footnote{The Aircraft Communications Addressing and Reporting System (ACARS)
will be superseded by the Aeronautical Telecommunications Network
(ATN) and IP communication over the next decade.%
} considerably limits its usefulness. It is also violating the basic
requirements of ADS-B as it is a system that sends on demand instead
of periodically broadcasting without being requested. Connection-oriented
protocols lack most of the advantages that the paradigm change with
ADS-B provides to modern ATC (specifically cost, scalability and ease
of use), which is why ADS-C has not been considered any further in
the development of NextGen.

\subsection{Communication in Military Avionics \label{sub:Communication-in-Military}}

There is undoubtedly a much stronger need and motivation to implement
stringent security in a military communications context. Though it
is not the primary focus of this survey, anything in practical use
by military forces could naturally be of interest for civil security
solutions as well. There are various standards currently in use by
the US and NATO military, among them the cryptographically secured
Mode 4 and Mode 5 as defined in the NATO Standardization Agreement
(STANAG) 4193. Mode 4, which employs a 3-pulse reply to a challenge,
has been in use for decades and according to the forecasts of the
NATO Minimum Military Requirements is to be superseded by Mode 5\emph{
}in 2015 (initial operational capability) and 2020 (full operational
capability), respectively \cite{Wagner2009}\emph{. }

While the legacy Mode 4 indeed only allows airplanes to respond to
challenges, Mode 5 adopts the ADS-B broadcast capability, so participants
can announce their presence without a prior query, very useful in
identification, friend or foe \cite{Assessment1993a}. On the security
side, Mode 5 uses proprietary hardware and encryption algorithms with
a black key concept;%
\footnote{Black keys are safe to transmit since they are encrypted with an encryption
key.\emph{ }Red keys on the other hand are unencrypted and classified
as highly sensitive.%
} it furthermore offers time-of-day authentication, automatic switchovers
and a longer period \cite{Kenney2008}. The signal modulation is done
via spread spectrum and operation requires a platform identification
number (PIN). Mode 5 hardware is equipped with a unique identifier
that informs about national origin and the platform number.\emph{
}Mode 5 has two different levels: Level 1 is the interrogation response
mode, providing time, position and identification based on both GPS
and traditional means\emph{.} Level 2 is the broadcast mode and entirely
based on GPS. There are currently no further available details on
the security mechanisms including the applied cryptography of Mode
4 and 5 since this information is classified.

The ADS-B specification itself also mentions the message types/downlink
formats Military Extended Squitter (DF19) as well as Military Use
Only (DF22) without detailing them further, although it is known for
example that DF19 makes ample use of bursts instead of regular beacon
messages only \cite{DO209}. Despite incomplete or unavailable information
on the performance of Mode 5 compared to ADS-B, it is safe to say
that both cost, scalability and ease of use of the known aspects of
the system are prohibitive to widespread use in commercial ATC. Spread
spectrum techniques and cryptography could, however, be a part of
a future security approach to ADS-B and will be discussed in this
survey. 
\begin{figure}
\includegraphics[scale=0.38]{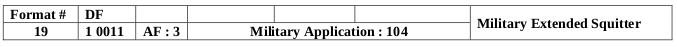}

\caption{DF19 data format \cite{DO209}}
\end{figure}

\section{Secure Broadcast Authentication \label{sec:Secure-Broadcast-Authentication}}

Secure Broadcast Authentication is one possible means to prevent and/or
detect attacks in a unidirectional broadcast network such as ADS-B.
This section will describe the various methods that have been proposed
in the literature, typically for wireless sensor networks or VANETs
and analyse their applicability to ADS-B. 

Authentication of messages on a broadcast medium is hard, compared
to point-to-point communication.\emph{ }A symmetric property is only
useful in point-to-point authentication where both parties trust each
other. Thus, an asymmetric mechanism is inherently required so that
receivers can \emph{verify} messages but are not able to \emph{generate}
authentic messages themselves\emph{ }\cite{Luk2006}. For a good overview
over secure broadcast communication in general, the reader is referred
to \cite{Perrig2003}.

The goal is to keep the open nature of ADS-B intact while offering
a potential authentication mechanism. This could be done either globally
or only selectively in cases where suspicious behaviour has been detected.
Such reactive authentication could lessen the strain on the network
by only requiring additional security (and thus computational and
communicational overhead) at times when incidents seem more likely. 

Furthermore, there is a distinction between broadcast schemes that
are user-based vs. those that are node-based, or possibly both. Node-based
(also known as host-based) schemes ensure the authenticity of a given
node, i.e. the hardware. User-based schemes on the other hand look
to authenticate a human user, regardless of the underlying hardware
\cite{Woo1992}. This survey focuses mainly on node-based schemes.\\

\subsection{Non-Cryptographic Schemes on the Physical Layer}

Non-cryptographic schemes such as fingerprinting comprise various
methods for wireless user authentication and device identification
techniques, either based on hardware or software imperfections or
characteristics of the wireless channel which are hard to replicate.
The goal is to identify suspicious activity in a network. Finding
a signature for legitimate beacons in a network, possibly being able
to tell apart ground stations from aircraft, identifying the type
of aircraft or even individual machines provides data useful for the
development of an intrusion detection system \cite{Danev2012}. If
there are tangible differences between legitimate and non-legitimate
packets on the physical layer, then machine learning techniques could
be employed to develop a model for predictions of normal behaviour
and also statistical thresholds beyond which an activity is considered
suspicious. Even if it is only feasible to identify classes of devices
instead of singular participants, this could prove to be valuable
information in detecting intruders. Yet, fingerprinting does not provide
surefire security in any way, and various attacks and concerns have
been brought forward \cite{Danev2010}.

Currently, there have been no attempts at applying any kind of non-cryptographic
schemes to boost the security of ADS-B. A common counter-argument
has been the fact that contrary to e.g. the 802.11 markets the commercial
airplane market is divided into two big players (Boeing and Airbus)
which in the long run makes significant differences at least between
ADS-B vendors unlikely. Still, fingerprinting has also been successfully
employed to tell apart the exact same models from the same vendor.
For a good overview of the state of the art in physical-layer identification
of wireless devices, see \cite{Danev2012}. 

Zeng et al.~\cite{Zeng2010} broadly identified three different techniques
that can be employed to enhance or even replace traditional cryptographic
measures: Software, hardware and channel-based fingerprinting.\\

\paragraph{Software-Based Fingerprinting}

This type of fingerprinting techniques tries to exploit distinctly
different patterns or behaviour of software operating on wireless
equipment. Depending on the specification of a protocol, there is
a lot of leeway for manufacturers and developers when implementing
software on a given device. If there is enough entropy in information
about the combination of chip sets, firmware, drivers to tell apart
different wireless users, this approach can be used to verify their
continuity up to a certain degree. As a downside, it seems likely
that large fleets of airline operators are fitted with very similar
or same hardware, making them harder or even impossible to differentiate
and on the other hand easier to study and copy for a potential attacker.\\

\paragraph{Hardware-Based Fingerprinting}

A number of techniques have been proposed to identify devices based
on unique hardware differences. Some of these differences can be used
for \emph{radiometric fingerprinting}, exploiting differences in the
turn-on/off transient (see, e.g.,~\cite{hall2005radio}) or the modulation
of a radio signal to build unique signatures. While this works well
for non-mobile cases and attackers with standard, off-the-shelf hardware,
it can break down against more powerful adversaries employing software
defined radios and be subjected to signal/feature replay attacks \cite{Danev2012}.
Furthermore, the existing research captured signals very closely to
the fingerprinting antenna (15\,m or less) and in non-mobile settings,
making it very improbable to work in the highly-dynamic, large-distance
ADS-B setting.

Another unique hardware feature amongst wireless devices is \emph{clock
skew}. As no two clocks run precisely the same, this can be used to
create signatures and enable identification. Unfortunately, to exploit
this, we would require timestamps included in ADS-B messages. Also,
it is possible for an attacker to eavesdrop on the communication and
mimic the appropriate clock skew \cite{Jana2010}.

Recently reviewed options for future systems include the use of so-called
physically unclonable functions (PUFs), which essentially exploit
specifically implemented circuits to create unique and secure signatures,
thus abandoning the scope of non-cryptographic solutions.%
\footnote{For a good overview on PUFs, see \cite{Maes2010}%
} Furthermore, besides requiring new hardware, this approach also necessitates
an overhauled messaging protocol, including a challenge and response
model \cite{Devadas2008}, making it a difficult fit for the requirements
of the ADS-B protocol.\\

\paragraph{Channel/Location-Based Fingerprinting}

Exploiting natural characteristics of the physical layer has been
a hot research topic in relation with security in wireless networks.
Various approaches have shown that this can be a viable alternative
to more traditional authentication and verification measures, typically
based on received signal strength (RSS, e.g.\,\cite{Mathur2008}),
channel impulse response (CIR, e.g.\,\cite{Zhang2010a}) or the carrier
phase (e.g.\,\cite{Wang2011a}). They are comparably easy to implement
in wireless systems and can offer reasonable security without requiring
much overhead.

Any such concept requires bidirectional communication, however. One
practical example is the retroactive authentication of data packets
via an RSS list as proposed by Zeng et al.~\cite{Zeng2010}. As this
temporal RSS variation authentication (TRVA) requires an ACK packet
in a given coherence time, it is not compatible with current ADS-B
protocols. Furthermore, it is doubtful if this could work with reasonable
efficiency in a highly dynamic environment such as airborne MANETs.
The coherence time $T_{C}$ in which the channel stays stable in a
wireless network where only one sender moves at 800km/h is roughly
0.6188ms. At a 1090ES bandwidth of 1 symbol/$\mu s$ it is obvious
that such protocols are impossible to deploy. This physical property
also effectively denies the application of many other more sophisticated
physical-layer schemes such as SecureAngle \cite{Xiong}, which aims
at securing wireless networks by using multiple antennas capturing
the angle of arrival information of nodes to built signatures and
detect anomalies. Similarly, practical indoor location-based geo-tags,
built with surrounding radio frequency signals such as done in \cite{Qiu}
are not applicable.%
\footnote{Although one advantage is that many legacy Mode S systems use several
directional rotating antennas to pick up ADS-B signals. The extracted
information about the angle-of-arrival could conceivably be used to
raise red flags, as described in Section \ref{sub:Traffic-Modeling-and-Plausibility-Checks}.%
}

Laurendeau and Barbeau \cite{Laurendeau2008,Laurendeau2009} exploit
RSS in a way similar to time difference of arrival concepts (see \ref{sub:Multilateration})
to localize malicious insiders in a vehicular ad hoc network with
the help of various receivers. Despite the fact that an attacker will
not be inclined to cooperate and can even actively fake the signal
strength he utilizes, their proposition enables the receivers of a
message to at least identify a given area where it must have originated
from.\\

\paragraph{Randomized/Uncoordinated Frequency Hopping / Spreading}

A physical-layer scheme different from fingerprinting, Frequency Hopping
Spread Spectrum (FHSS) and Direct Sequence Spread Spectrum (DSSS)
are both used in wireless systems to improve protection against malicious
narrow band and pulse jamming as well as eavesdropping. In their usual
form they both require a pre-shared spreading code or hopping pattern
between sender and receiver which makes it hard to follow or hinder
the communication for anyone without access to the code/pattern. This
is also exploited in military communications (see Section \ref{sub:Communication-in-Military})
but is not a viable option for world-wide civil and commercial ATC
where such secret codes would presumably not stay secret for long.

The need for a pre-established code can be relinquished by employing
random, uncoordinated versions of FHSS and DSSS. Strasser et al.~\cite{Strasser2008}
propose such a physical layer approach to counteract jamming in wireless
broadcast scenarios. Uncoordinated Frequency Hopping (UFH) provides
a viable way to broadcast initial messages without an attacker being
able to jam the transmission in an efficient way. The key insight
to these approaches is that, contrary to normal frequency hopping
mechanisms, sender and receiver(s) rely on the statistical chance
to be on the same channel at the same time. 
\begin{figure}
\includegraphics[scale=0.5]{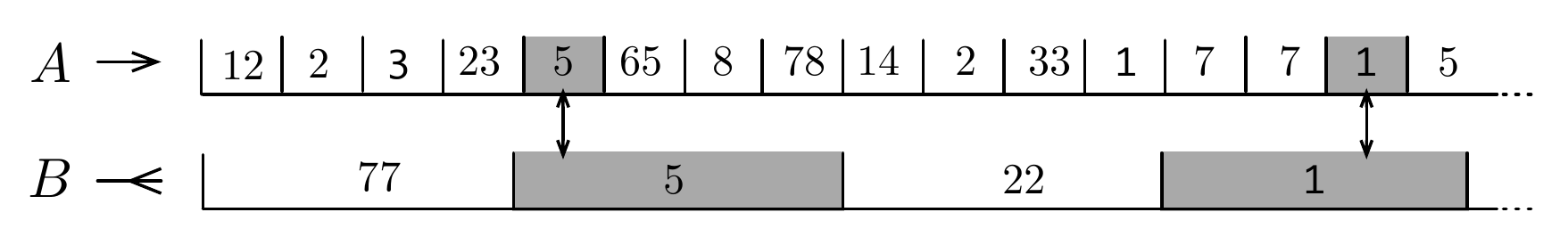}

\caption{Uncoordinated Frequency Hopping after \cite{Strasser2008}. Both A
and B regularly change their communication frequencies without having
pre-established a common pattern. By statistical chance they will
communicate on the same channel every so often. }

\end{figure}
The obvious downside of UFH is its low bandwidth due to the fact that
many times receivers will not listen on the correct channel. More
concretely, the probability with UFH that a packet will be received
at a node without an attacker being involved is $p_{m}\geq1-(1-\frac{c_{s}}{c})^{c_{r}}$
($c$ being the number of possible channels, $c_{s}$ and $c_{r}$
the number of channels a sender/receiver is using simultaneously).

Uncoordinated Direct-Sequence Spread Spectrum (UDSSS) \cite{Christina2009}
and Randomized Differential DSSS \cite{Liu2010} are techniques based
on the same principle. They rely on the statistical chance that spread
codes randomly chosen by sender and receiver(s) will happen to be
the same every so often. 

While the proposed methods can effectively defeat jamming and modification
attacks, the inherently lower performance and a prolonged transmission
time make them difficult to use in a large-scale system such as ADS-B.
Furthermore, authentication and security against replay attacks is
only achieved by adding a private/public key infrastructure and timestamps,
respectively.\\

\subsection{Public Key Cryptography}

Cryptographic measures have been a tried and tested means to secure
communication in wireless networks and must subsequently also be considered
in the ADS-B setting. One question to examine is if the current implementation
of ADS-B can be encrypted. The first possibility would be to distribute
the same encryption keys to all ADS-B participants worldwide, or at
least to aircraft and ground stations in a given area. Such a vast
group encryption scheme, including even general aviation, would be
considered extremely insecure to both inside and outside attacks.
This inherent weakness is non-fixable even with very frequent key
updates (which also would again increase the complexity of the encryption
deployment). In short, such a scheme would fulfill none of the required
criteria listed in Section \ref{sub:Requirements}. Finke et al.~\cite{Finke2013a}
examine various encryption schemes, including the possibility to do
the key management for symmetric encryption out of band, for example
through the controller-pilot data link communication (CPDLC) which
they consider worth exploring further. The authors also give an analysis
of the security and practicability of asymmetric, symmetric and format
preserving encryption. In their conclusion, they support a symmetric
cipher using the FFX algorithm (format-preserving, Feistel-based encryption
with multiple implementation variances) which can encrypt non-standard
block sizes (i.e.~ADS-B's 112 bit messages) with sufficient entropy.
However, the difficulties concerning key management and distribution
are strongly acknowledged. Most recently, Wesson et al.~\cite{wessoncan}
look at the broader question of how to use encryption to secure ADS-B
and conclude that the problems with symmetric cryptography are too
large to overcome. They argue that PKI is the only feasible cryptographic
approach and propose ECDSA signatures as the smallest and thus best
solution. Besides the key management problem, they further analyse
the interference burden on the ADS-B channel, showing that even without
the significant additional traffic that is currently found on the
1090\,MHz frequency, the decrease in operational capacity would potentially
be crippling.

As mentioned before, if broadcast authentication is needed, one requires
an asymmetric property, a characteristic fulfilled by public key cryptography.
Samuelson and Valovage \cite{Samuelson2006} report on an implementation
of authentication and encryption in UAT using a public key infrastructure
(PKI). Their method uses a hash to create a message authentication
code (MAC) that can be used to authenticate the message and can be
extended to full encryption but no further details are publicly available.
There are related patents filed under the names ``Secure ADS-B Authentication
System and Method'' \cite{viggiano2010secure} and ``Automatic Dependent
Surveillance System Secure ADS-S'' \cite{schuchman2011automatic}
by Sensis Corporation.%
\footnote{Now owned by defense contractor Saab AB.%
} The general idea is to use a challenge/response format with an authenticator
ground station, who authenticates every participant in its reach and
notifies a higher authority and/or all other participants of any failed
authentication. This concept requires the station to have access to
a worldwide database of secure keys that is both hard to maintain
globally as well as subject to possible security breaches. If this
is the case, the system can be used to not only identify ADS-B participants
but also pilots and various other people/systems taking part in the
flight process. Furthermore, ADS-S includes a changed modulation and
a complete overhaul of the messaging system, making it incompatible
with ADS-B and as such very costly and extremely doubtful to be deployed
widely in the future.
\begin{figure}
\includegraphics[bb=0bp 0bp 730bp 777bp,clip,scale=0.375]{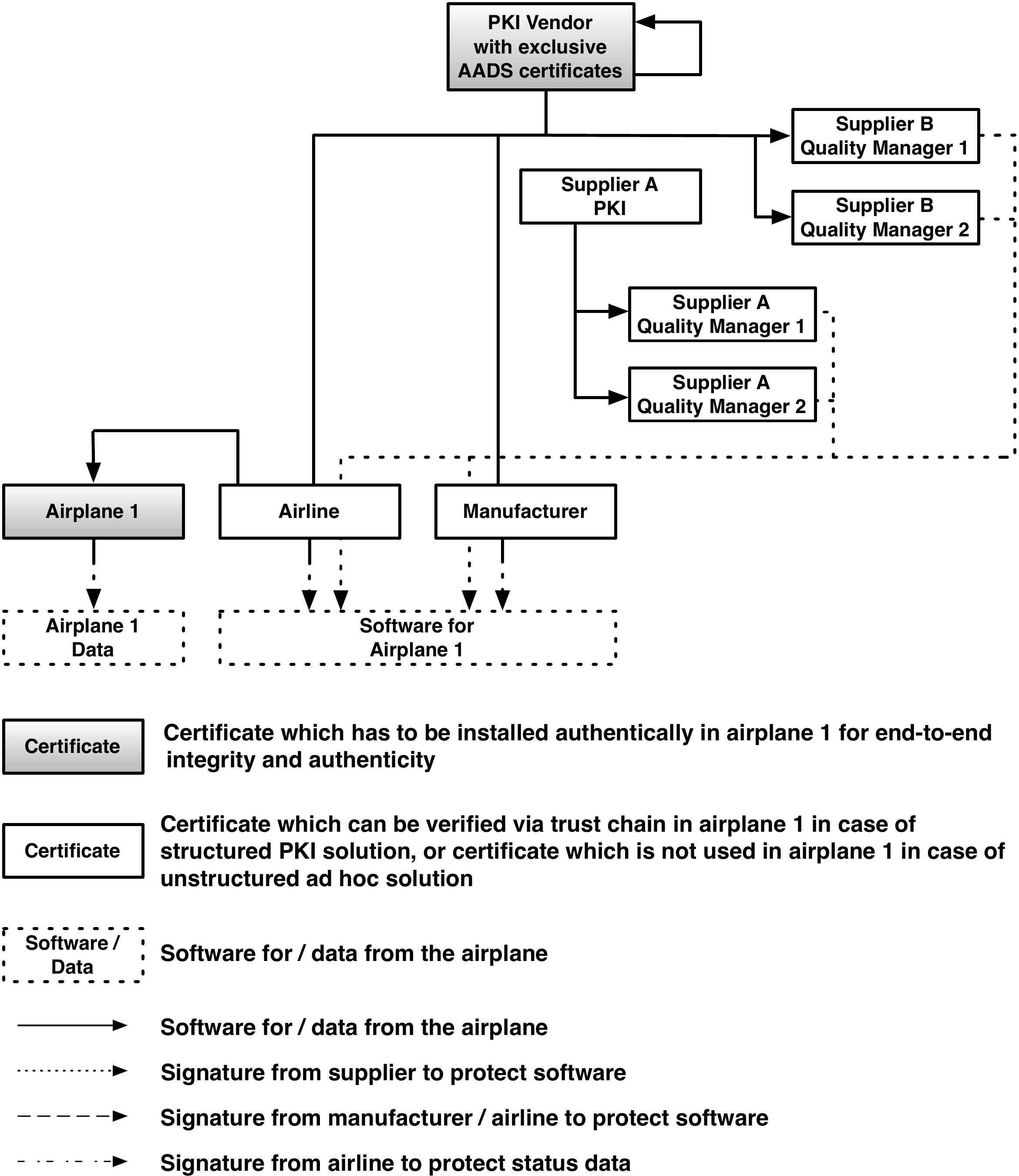}\caption{Outline of a potential public key infrastructure after \cite{Robinson2007}.
A PKI vendor supplies airlines, manufacturers and external suppliers
with the required certificates. \label{fig:Outline-of-aPKI} }
\end{figure}

Costin et al.~\cite{Costin} suggest a ``lightweight'' PKI solution
which essentially amounts to a retroactive part publication of the
key as discussed in Section \ref{sub:Retroactive-Key-Publication}:
Aircraft $A$ transmits the signature distributed over a number $N$
of ADS-B messages, so that after every $N$ messages the surrounding
participants have received $A$'s signature. The recipients keep the
messages until the full signature has been transmitted, at which point
they can authenticate the buffered messages.\emph{ }The authors suggest
that the PKI key distribution necessary for this scheme could be done
during an aircraft's regular check-ups. 

Ziliang et al.~\cite{Ziliang2010} present a concrete PKI solution
for data authentication in ADS-B/UAT based on Elliptic Curve Cipher
and X.509 certificates. The authors try to tackle a number of problems
involved with cryptography such as key size length and keeping the
broadcast nature of ADS-B. UAT offers much longer messages than 1090ES,
with payloads of 16 or 32 bytes when transmitted from aircraft and
even 464 bytes in ground message bursts \cite{Moody2000}. Yet, their
conclusion is that the data block format needs to be changed, no matter
which type of cryptography is used. Consequently, they propose and
implement their authentication scheme with a slightly changed UAT
message type. In 1090ES, this scheme would require not only using
the DF24 Extended Length Messages available in the Mode S standard
but still need 5 messages to divide and accommodate the signature
data and timestamps, a solution that hardly looks scalable in an already
crowded frequency. On top of this, the description leaves very much
open the question of an efficient certificate distribution scheme.

Raya and Hubaux \cite{Raya2005,Raya2007} discuss using Public Key
Cryptography in VANETs. The considered scenarios were short-range
with beacons sent every 100-300ms and included up to 120 mobile nodes
in a 300m communication range. They looked at message sizes between
294 and 791 bytes and found the performance to be acceptable in their
simulations.

Robinson et al.~\cite{Robinson2007} analyse various different solutions
to create PKI infrastructures for a general airplane assets distribution
system (AADS). Although, the work is not discussing the ADS-B protocol
but instead focuses on the distribution of software and data on the
ground, the authors identify the airline industry's needs and requirements
from a PKI infrastructure, and it seems plausible that the same system
could be used to secure air traffic control data. According to their
analysis, an ad hoc approach without a central authority, employing
pre-loaded trust certificates, could be used as a short-term solution
until a more structured, long-term public key infrastructure has been
developed (see Fig.\,\ref{fig:Outline-of-aPKI} for an outline of
their proposal).

The obvious idea for a centralised key distribution would be to have
aviation authorities such as the FAA act as a certificate authority
(CA).\emph{ }But assuming the role of a CA is no easy task. Even many
specialized institutions had to report numerous security breaches
over the last decades. Furthermore, if this problem is sufficiently
solved, there remains the question of how aircraft from airspaces
mandated by different authorities can securely communicate with each
other. These challenges are somewhat analog to the same approach in
vehicular networks as discussed in \cite{Parno} but arguable even
worse due to the large internationalization of the ADS-B network.\\

\begin{figure}
\includegraphics[scale=0.35]{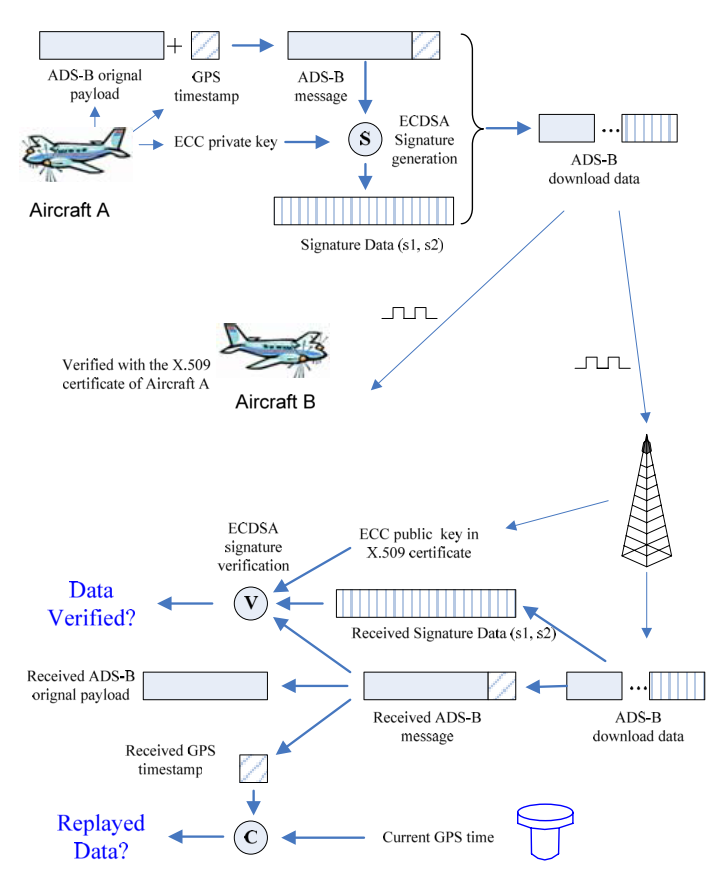}

\caption{Example of a typical encryption scheme adapted for ADS-B from \cite{Ziliang2010}.
It employs elliptic curve cryptography to generate signatures which
can be verified with the responding certificate by other aircraft
and ground stations. An additional GPS timestamp prevents replay attacks.}
\end{figure}

There are certain natural disadvantages to using an encryption solution
that cannot be overcome (or only with great difficulty) as mentioned
in \cite{Zhang2010}: 
\begin{itemize}
\item Despite the encryption of data frames, management and control frames
are not protected.
\item It immediately and unmitigatably breaks compatibility with the installed
base. 
\item Key exchange is notoriously difficult in ad hoc networks, which are
by definition without a centralized institution. They are often too
dynamic, requiring constant adaptation. This would result in too much
overhead in both the number and the size of messages. 
\item The open nature of ADS-B is widely seen as a feature. A cryptographic
system implemented in a way comparable to ADS-S does not offer public
broadcast communication.
\item One-time signatures even using advanced techniques such as Merkle-Winternitz
prove infeasible due to their overhead of 80 bytes and more, simply
to sign 60 bits \cite{Luk2006}.
\end{itemize}
To conclude this section, it is difficult to build any kind of encryption
scheme with the currently standardized 1090ES data link. Approaches
have been shown to be theoretically possible with the higher bandwidth
UAT, although practical proof of scalability and practicability have
not been given yet. Furthermore, at this point in time it does not
seem likely that UAT will play a role apart from general aviation
in the FAA-mandated airspace. So, while UAT offers more technical
possibilities not only for security and encryption, and even combined
UAT/1090ES transmitters are neither a technical nor a regulatory problem,%
\footnote{``There is nothing in the regulations and there are no technical
hurdles that would prevent a manufacturer from building a combination
UAT/1090ES box, one that would eliminate the different-technology
blind spots while allowing high-flying aircraft the ability to get
FIS-B through UAT.'' http://www.flyingmag.com/technique/proficiency/ins-and-outs-ads-b%
} traditional cryptography in conjunction with the current installment
of ADS-B seems to be a very difficult route for further research at
the present.

\subsection{Retroactive Key Publication\label{sub:Retroactive-Key-Publication}}

A variation on traditional asymmetric cryptography is the technique
of having senders retroactively publish their keys which are then
used by receivers to authenticate the broadcast messages. This approach
that has been proposed for use in various fields \cite{Perrig2003,Perrig2000}.
The key concept is simple: Any broadcasting entity produces an encrypted
message authentication code (MAC) which is then sent along with every
message. After a set amount of time or messages, the key to decrypt
this MAC is published. All listening receivers, who have buffered
the previous messages, can now decrypt the messages and ensure the
continuity of the sender over time.

The TESLA (Timed Efficient Stream Loss-Tolerant Authentication) protocol
\cite{Perrig2005}, standardized in RFC 4082,%
\footnote{``Timed Efficient Stream Loss-Tolerant Authentication (TESLA): Multicast
Source Authentication Transform Introduction'', http://www.ietf.org/rfc/rfc4082.txt%
} can provide efficient broadcast authentication on a large scale,\emph{
}while it is able to cope with packet loss and real-time applications\emph{.
}The μTESLA broadcast authentication protocol is the adaptation of
TESLA for wireless sensor networks \cite{Perrig2002}.

\begin{figure}
\includegraphics[scale=0.45]{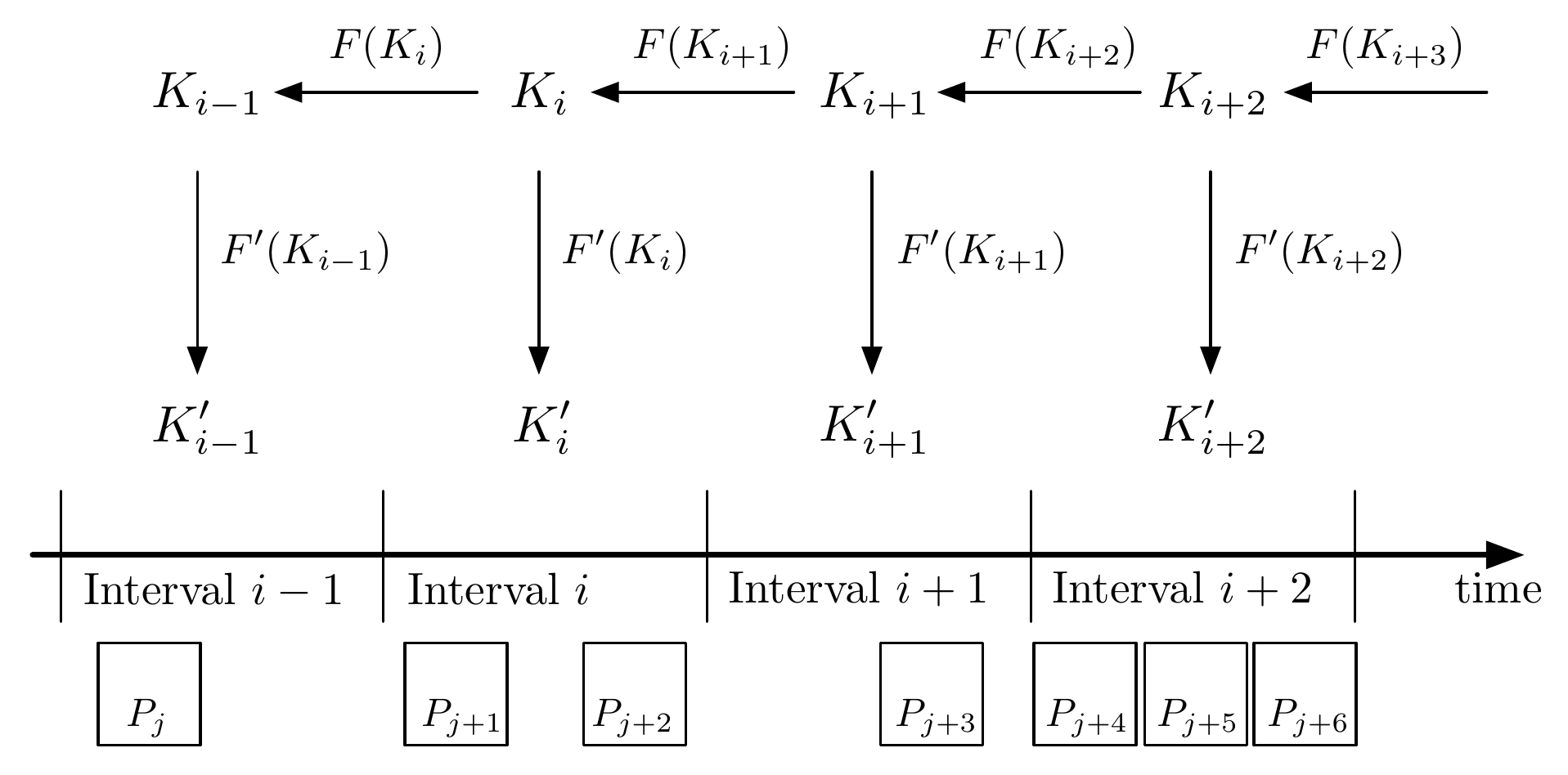}

\caption{The figure illustrates TESLA's utilization of one-way chains after
\cite{Perrig2005}. The first one-way function $F$ generates the
chain, following that the second one-way function $F'$ derives the
MAC keys. Time is divided into separate intervals $i$, all having
the same length. The packets $P_{j}$ are each sent during one specific
interval. For every such packet, the sender computes a MAC with the
key that is in accordance with that interval. E.g.\emph{ $P_{j+2}$}'s
MAC is calculated based on its data and key $K'_{i}$. Disclosing
the keys of previous intervals can be done either by attaching the
key to sent packets or in separate messages. \label{fig:TESLA}}

\end{figure}

Both TESLA and μTESLA use one-way key chains as shown in Fig.\,\ref{fig:TESLA}:\emph{
}The broadcaster chooses a random key $K_{n}$ and applies a public
pseudo-random function $F$\emph{ }as often as required to acquire
the keys: \emph{$K_{i}=F(K_{i+1}),\,0\leq i\leq n-1$.} Subsequently,
every secret $K_{i},i>0$ is used for sending in the $i$-th interval
and disclosed to the public after a number of time intervals $d$.
As every previous key $K_{i}$ with $i<j$ can be recovered by the
receiver(s) by applying the one-way function $F$, the receiver needs
to do two things to authenticate a message: \cite{Liu}
\begin{enumerate}
\item Authenticate the key $K_{i}$ against previously received keys to
ensure they are from the same key chain.
\item Ensure that the message with key $K_{i}$ could only have been sent
before the key has been published (requiring loose time synchronization),
i.e. before interval $i+d$.
\end{enumerate}

The fact that μTESLA uses symmetric cryptography in connection with
time as its asymmetric property makes it an interesting idea for adaptation
to ADS-B since a sufficiently good time synchronization could be provided
via GPS (this would require sending the GPS timestamps in a new protocol
field since this is not currently the case). The advantages of μTESLA
are obvious: ADS-B keeps its open and broadcast nature and a complex
PKI infrastructure is not required to ensure a sender's continuity,
although it could be added if identification and source integrity
are required (e.g. solely for well-connected ground stations). Nonetheless,
it enables a participant to protect itself against impersonation attacks.
In areas well-served by ground stations any break in continuity detected
by any single one of them would set off red flags.

Another advantage of μTESLA is that lost packets on the notoriously
jammed 1090 MHz frequency (there is no medium access control in place)
are not an integral problem for authentication. Furthermore, the overhead
for communication as well as required modifications to the ADS-B protocol
are significantly less than with traditional asymmetric cryptographic
methods.%
\footnote{ The original μTESLA requires a 6 byte MAC.%
}

\begin{figure}
\includegraphics[scale=0.33]{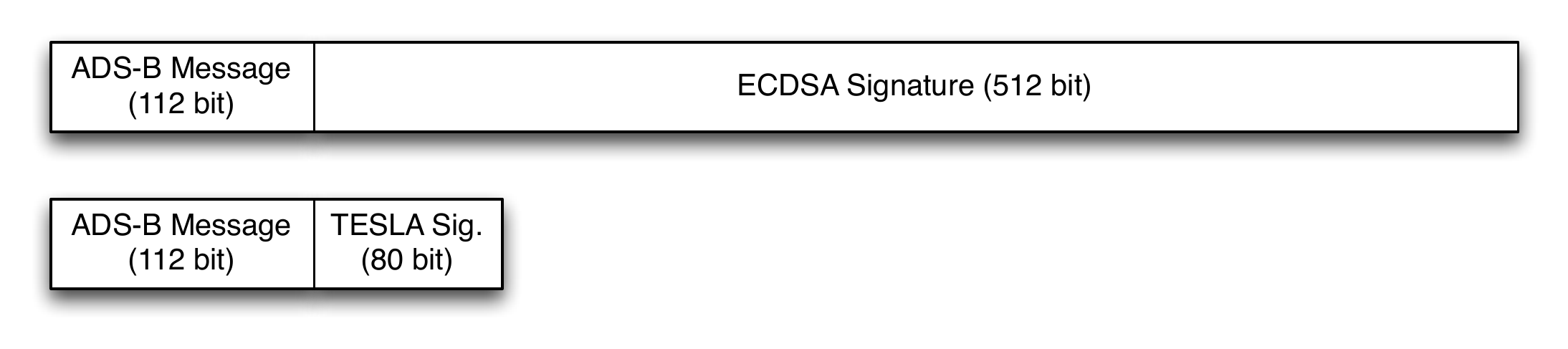}

\caption{TESLA signatures vs.~ECDSA signatures. Tesla signatures cost significantly
less overhead compared to many other cryptographic solutions. \label{fig:TESLA-signatures-vs.}}

\end{figure}

On the other hand, μTESLA also has some disadvantages when applying
it to AANETs.\emph{ }There is a need to reinitialize it (if used for
identification) and it can be susceptible to memory-based DoS (depending
on the setup of the receiver).\emph{ }To counter this, Eldefrawy et
al.~\cite{Eldefrawy2010}\emph{ }propose an approach that utilizes
forward hashing using two different nested hashes and the Chinese
Reminder Theorem,\emph{ }resulting in a system that does not need
to be reinitialized. 

Haas and Yu \cite{Haas2009} compare TESLA and ECDSA-based authentication
by simulating their performance in a real-world VANET scenario (although
not including certificate distribution). Regarding channel congestion
and MAC layer delay, they found that the TESLA protocol with keys
attached to a subsequent broadcast performed significantly better
than ECDSA or a TESLA-scheme that publishes keys in separate packets.

Hu and Laberteaux \cite{Hu2006} discuss a combination of a full-blown
PKI infrastructure with CAs distributing 512 bit signatures to bootstrap
80 bit TESLA signatures for short-term authentication in VANETs. Such
a system offers comparably lightweight integrity and possibly on-demand
authentication, if needed and requested. \\

\section{Secure Location Verification \label{sec:Secure-Location-Verification}}

Besides securing the communication - and thus the location data -
of ADS-B, there are other approaches to ensure the integrity of air
traffic management. The general idea of secure location verification
is to double check the authenticity of location claims made by aircraft
and other ADS-B participants. This is inherently different from the
verification of the broadcast sources and messages. The baseline is
to establish means to find the precise location of a sender, effectively
offering some redundancy and thus the ability to double check any
claims made. As an additional advantage any such approach creates
more location data, which can be merged with ADS-B and radar and offer
a back-up system in case of failure of these primary navigation systems
or GPS.

\subsection{Multilateration\label{sub:Multilateration}}

Multilateration, or hyperbolic positioning, is a popular form of \emph{co-operative
independent surveillance} and has been successfully employed for decades
in military and civil applications. If the precise distance between
four or more known locations and an unidentified location can be established,
it is a purely geometric task to find the unknown point. We can, for
example, use the received ADS-B signals which travel at the speed
of light to estimate the distance. Since we do not know the absolute
time a message needed to travel from an airplane to a receiver, we
have to employ the time difference of arrival (TDOA).%
\footnote{For a full explanation of the multilateration process in aviation
see e.g.\,\cite{Savvides2002} or \cite{Neven2005}. For an overview
over wireless location techniques, see \cite{sayed2005network}.%
}\emph{ }

\begin{figure}
\includegraphics[scale=0.4]{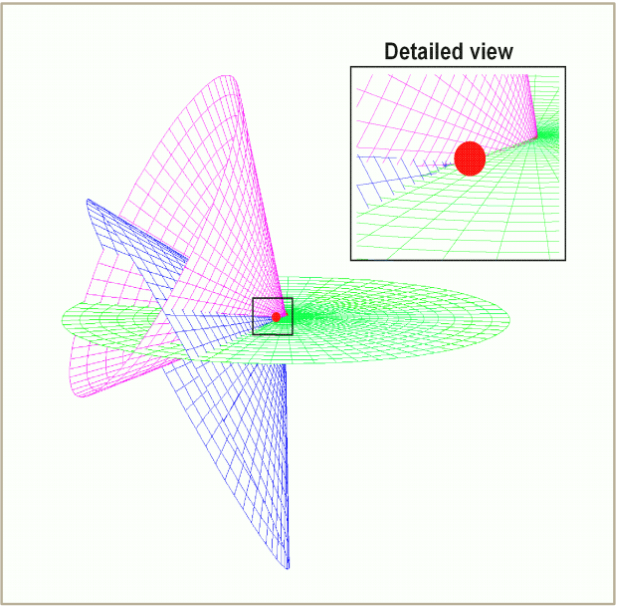}

\caption{Intersection of three hyperboloids from \cite{Xu2010}. With four
receivers in a 3D setting, one can specify the origin of the message
as the red point where the computed hyperboloids intersect. \label{fig:Intersection-of-three}}

\end{figure}

Thus, multilateration requires a number of antennas\emph{ }in different
locations that receive the same signal at different times. From the
TDOA, hyperboloids can be calculated on which the aircraft's position
must lie.\emph{ }With four or more receivers, a 3D position can be
estimated by finding the intersection of the hyperbolas as shown in
Fig.\,\ref{fig:Intersection-of-three}.

\begin{figure}
\begin{centering}
\includegraphics[scale=0.5]{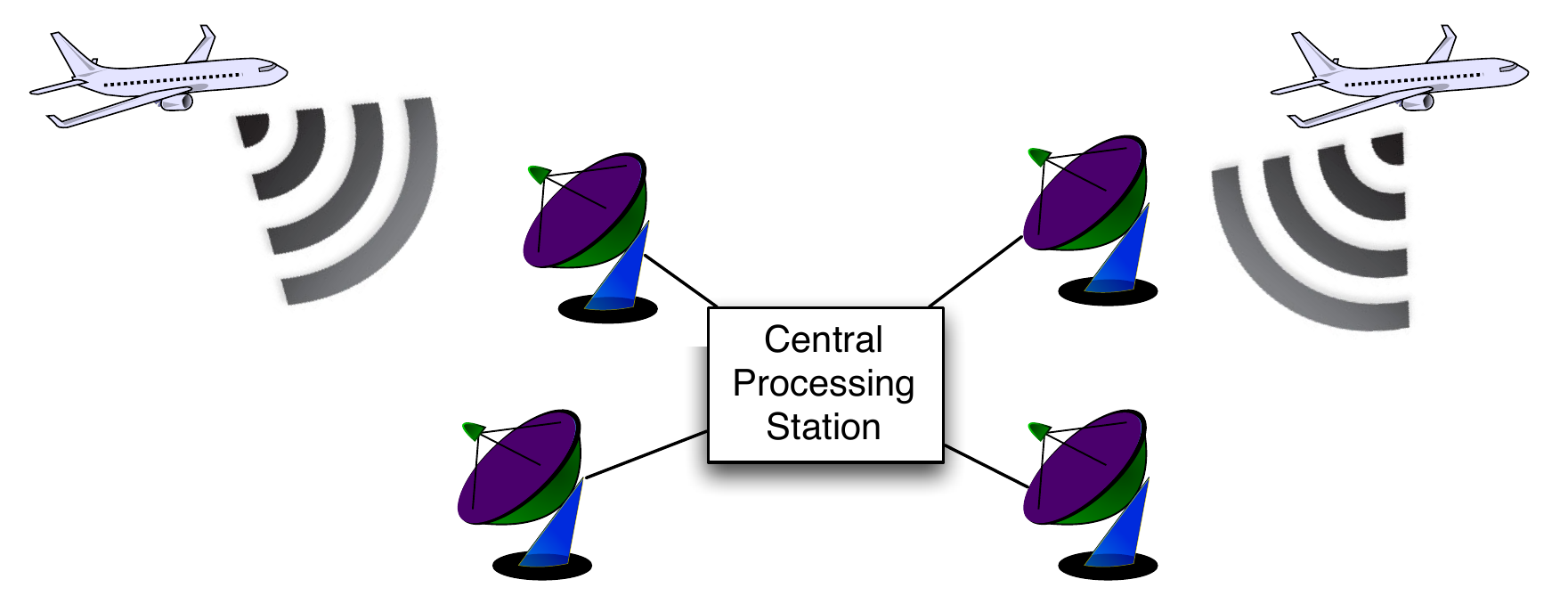}
\par\end{centering}

\caption{Basic multilateration architecture. Four (or more) receiver stations
measure the time at which they receive the same message from an aircraft.
They send this data to the central processing station which can calculate
the aircraft's position from the time difference of arrival between
the receiver stations. \label{fig:MLAT-architecture.}}
\end{figure}

Performing multilateration by utilizing TDOA is currently the preferred
solution for location verification on the ground. It is used in the
field (e.g. by the ASDE-X system \cite{HerreroJ.G.;BesadaPortasJ.A.;RodriguezF.J.Jimenez;Corredera1999})
at various US airports%
\footnote{See the manufacturer's description: http://www.saabsensis.com/docs/128/%
} and also being rolled out in Europe in connection with the CASCADE
project.%
\footnote{http://www.cascade-eu.org%
} One major advantage of multilateration is the fact that it can utilize
aircraft communication that is already in place. Thus, there are no
changes required to the currently existing infrastructure in aircraft,
while on the ground receiver stations and central processing stations
have to be deployed (see Fig.\,\ref{fig:MLAT-architecture.}).

\begin{figure}
\includegraphics[bb=5bp 5bp 838bp 463bp,clip,scale=0.3]{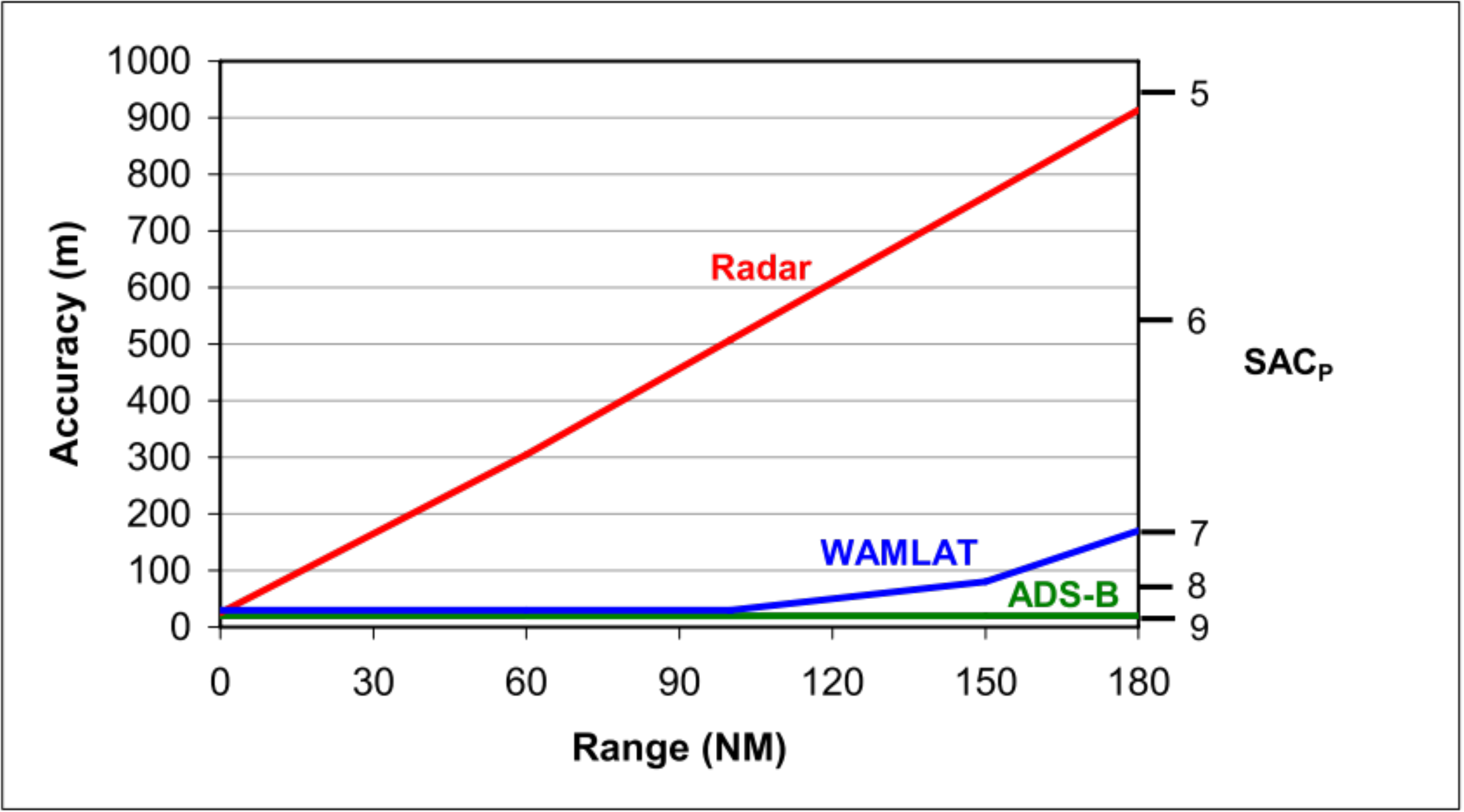}

\caption{Comparison of location estimation accuracies when utilizing primary
radar, wide area multilateration and ADS-B \cite{Smith2006}. \label{fig:Comparison-of-location}
$SAC_{p}$ denotes the Surveillance Accuracy Category for Position
as defined in \cite{Smith2006}.}
\end{figure}

While currently used mainly in comparably short distances (taxiway
and runway on airports, up to about 60\,m height), Wide Area Multilateration
(WAMLAT) has also been a popular research topic. Compared to primary
radar systems, WAMLAT is relatively easy and cost-effective to install
and use on the ground but can also be successfully employed in an
airborne MANET.%
\footnote{Although vast open spaces such as found in e.g. Australia or over
oceans can prove infeasible for the use of multilateration and have
been one of the very reasons driving the development and deployment
of ADS-B.%
} Using an estimated distance of a target between four or more receivers,
it is possible to tell the 3D position of a sender with roughly 30\,m
accuracy (at 90\,NM distance) compared to 20\,m for ADS-B \cite{Purton2010}.
However, in comparison to ADS-B the accuracy of multilateration in
practice deteriorates over long distances (see Fig.\,\ref{fig:Comparison-of-location}). 

There have been various practical studies of multilateration using
ADS-B signals. For example, \cite{Smith2006} examines it as a method
to provide a means to backup and validate ADS-B communication. Johnson
et al.~\cite{Johnson2012} describe their proof of concept work in
a war-zone in Afghanistan. Kaune et al.~\cite{Kaune2012} built a
proprietary low cost test bed to do multilateration with ADS-B signals.
Thomas \cite{Thomas2011a} presents findings from two controlled helicopter
flights in the North Sea.

Daskalakis and Martone \cite{Daskalakis2003} give a technical assessment
of the possibility of using ADS-B and WAMLAT in the Gulf of Mexico,
testing a single controlled flight with good accuracy. 

A recent work at MITRE Corporation \cite{Niles2012} analyzes the
attempt to build an alternative navigation system with WAMLAT in the
FAA-mandated US airspace. The authors discuss the potential use of
already deployed sensors for multilateration around three airports
with flat terrain, also noting that the challenge is greater in more
mountainous areas. Further, they provide some sensor placement discussion,
determining the optimal choice and number from a given database concerning
requirements such as accuracy and low dilution of precision. 

Despite its successful use in the field, multilateration leaves a
number of open problems in terms of secure location verification,
for example the estimation of aircraft altitudes with ground-based
receivers is known to be very difficult. Galati et al.~\cite{Galati2005}
discuss the theoretical application of SSR as the basis for multilateration
in airport surveillance. They analyze a case study of the Marco Polo
airport in Venice and look at technical details such as dilution of
precision and multilateration algorithms and conduct simulations with
five sensors in a 25\,km radius around the airport. They propose
angle-of-arrival measurements to improve the unsatisfying height estimates
provided by the wide area multilateration. 

The International Civil Aviation Organization also names a few known
drawbacks of multilateration: \cite{Siu2011}
\begin{enumerate}
\item It is susceptible to multi-path propagation.
\item A signal has to be correctly detected at comparably many receiving
stations.
\item A separate link between the central processing station and all receivers
is required. 
\end{enumerate}
Essentially, some of the limitations stem from cost and logistic reasons
as it can be difficult and expensive to deploy enough sensors and
stations in very remote and inaccessible areas. 

Furthermore, \cite{schuchman2011automatic} mentions an attack vector
for an adversary trying to fool a receiver system utilizing multilateration
for location verification. An attacker would need to purchase and
modify four traffic-collision avoidance system (TCAS) receivers\emph{
}and use a GPS/WAAS time transfer unit between the units\emph{ }to
ensure relative timing accuracy. Furthermore, he needs to engineer
an algorithm similar to the one TCAS uses to determine aircraft's
tracks. As this involves both a certain cost and non-zero engineering
knowledge the difficulty of exploiting this threat is relatively high,
certainly when compared with the simplicity of spoofing ADS-B messages. 

In principle, there always remains the question of how to secure the
communication needed to perform multilateration and relay the localisation
information to the other participants. If it is not properly secured,
Sybil attacks, where a number of dishonest nodes deceive their environment,
are entirely possible \cite{douceur2002sybil}. \\

\subsection{Distance Bounding\label{sub:Distance-Bounding}}

\begin{figure}
\begin{centering}
\includegraphics[scale=0.45]{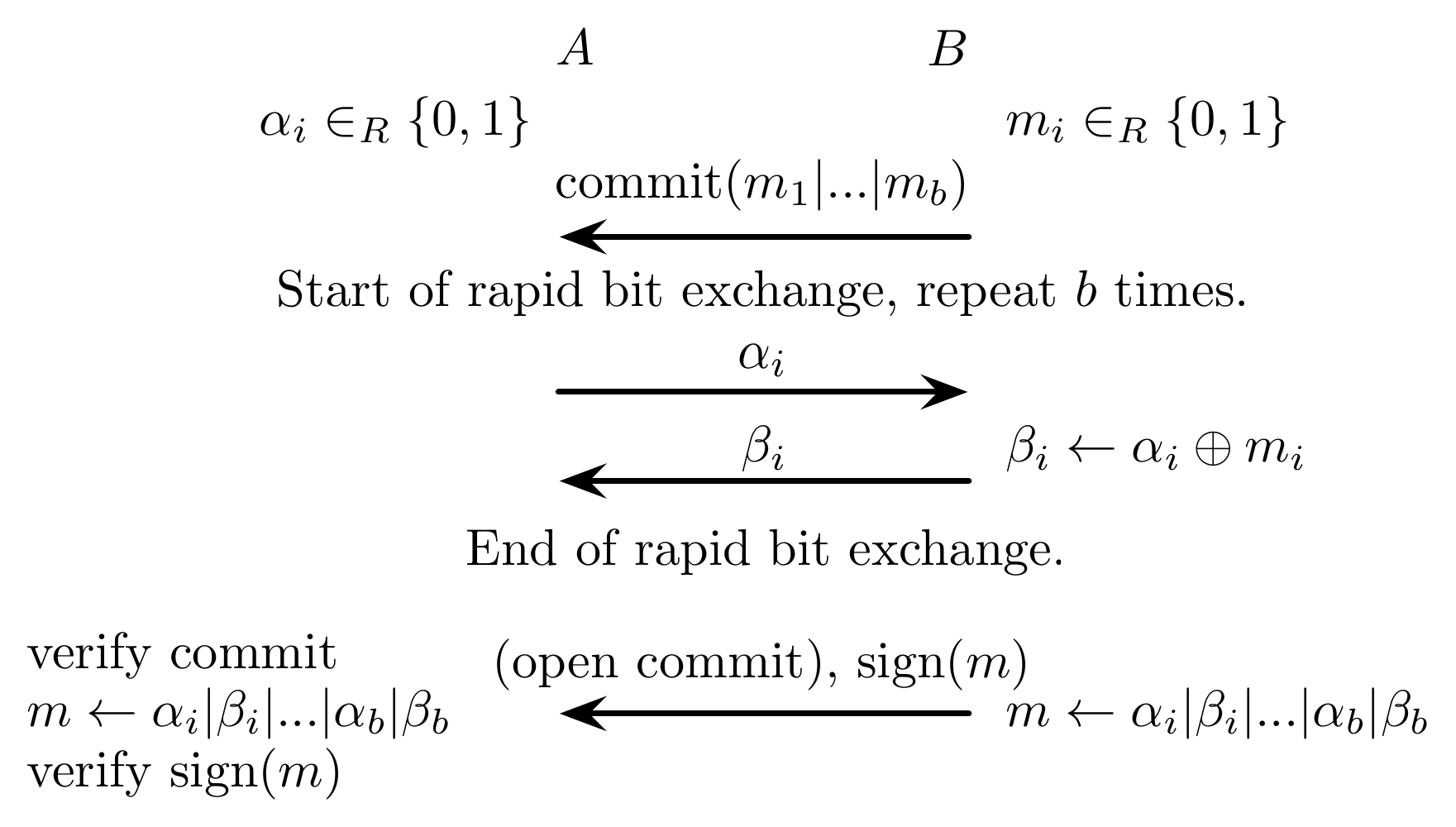}
\par\end{centering}

\caption{Original untrusting distance bounding protocol by Chaum and Brands
\label{fig:Original-untrusting-distanec} \cite{Brands1994}. $A$
is the verifier, $B$ the prover. After the protocol exchange $A$
can verify that $B$ is within a given distance.}

\end{figure}

Distance bounding is another method that has been employed in wireless
networks to partly localize other participants and ensure secure transactions
e.g. for RFID communication. First presented by Brands and Chaum in
1993 \cite{Brands1994}, the idea behind distance bounding is to establish
a\emph{ }cryptographic protocol with the goal to have a prover $P$
show to a verifier $V$ that $P$ is within a certain physical distance
(see Fig.\,\ref{fig:Original-untrusting-distanec} for the concrete
protocol).\emph{ }The universally valid fact that electro-magnetic
waves travel roughly at the speed of light \emph{c}, but never faster,
builds the foundation of all distance bounding protocols.%
\footnote{This is in contrast to e.g. distance estimation via received signal
strength, which can be influenced and faked by a malicious node.%
} This enables the computation of a distance based on the time of flight
between the verifier's challenge and the corresponding response by
the prover. 
\begin{figure}
\includegraphics[scale=0.47]{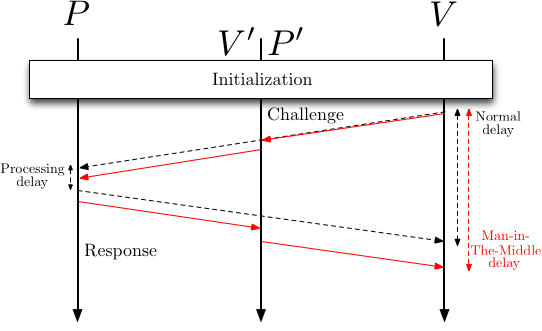}

\caption{Principle of distance bounding protocols. The verifier $V$ sends
a challenge to the prover $P$ who then, after processing, sends his
response (black dashed arrows). A man in the middle ($V'$/$P'$)
can only increase the distance by adding further processing delays,
but not decrease it (red arrows).}

\end{figure}
The determined distance serves as an upper-bound, an additional piece
of information that can subsequently be used as a means to verify
and authenticate a node by checking the truth of its claims. When
distance-bounding is performed by various trusted entities (such as
ground stations) these can collaborate and find the actual location
of the prover via trilateration.%
\footnote{Triangulation is not to be confused with the previously described
multilateration. The former uses the absolute measurements of three
or more distance circles to determine positions, while the latter
uses the \emph{difference} in distance between the measurements.%
} There are various practical attacks on distance bounding schemes
given in the literature, among them a number of relay attacks such
as the so-called distance fraud, mafia fraud and terrorist fraud \cite{Clulow2006}
as well as the newer distance hijacking attack \cite{Cremers2012}.

\begin{figure}
\includegraphics[bb=0bp 0bp 576bp 323bp,scale=0.44]{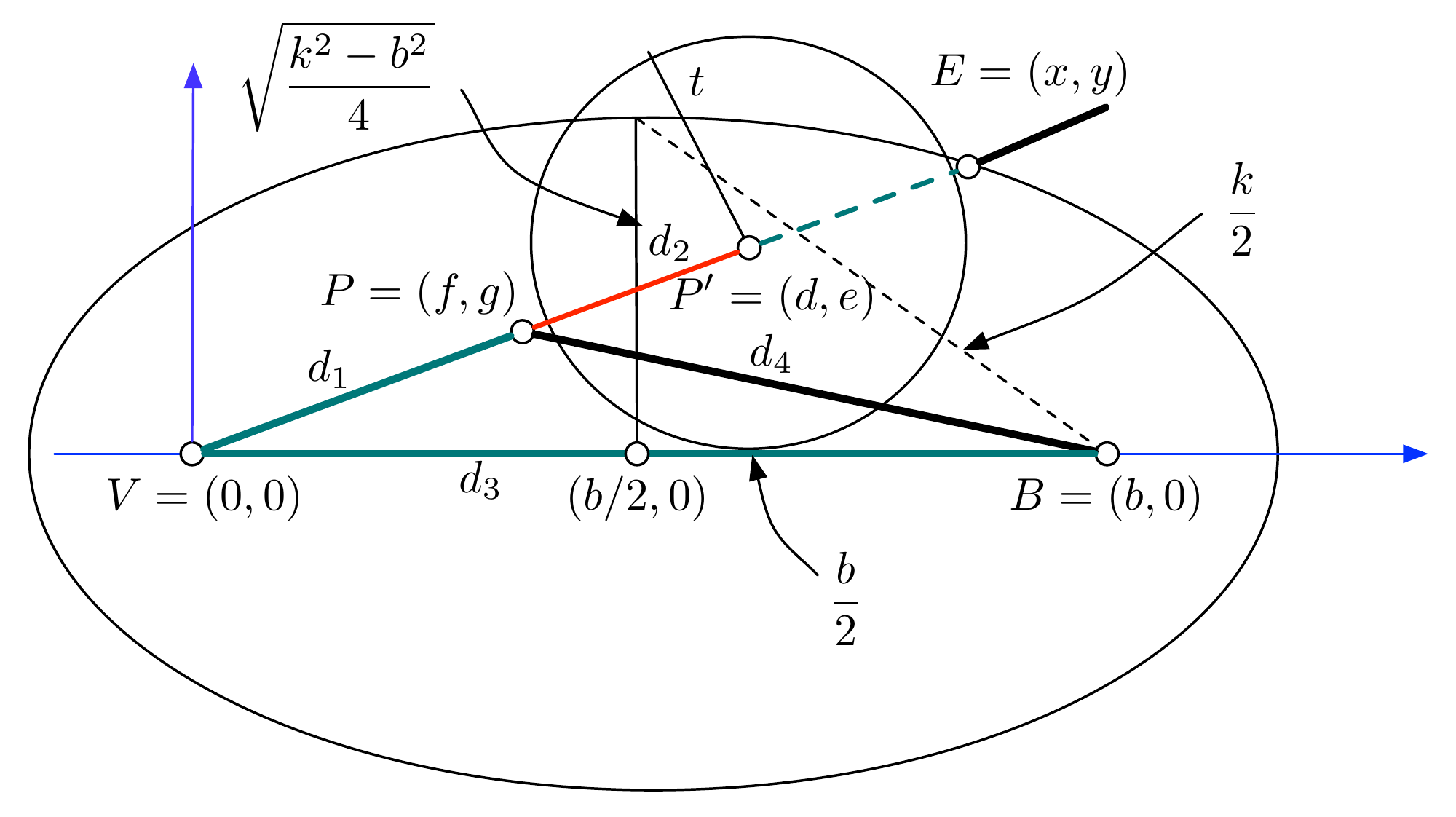}

\caption{Network topology for a minimum distance guarantee after \cite{Song2008}.
$V$ is the verifier, $P$ the prover, $P'$ its claimed location.
$B$ is a common neighbour of both $V$ and $P$ who gives an estimate
$E$ of $P$'s position. Considering the ellipse and a certain error
distance, $B$ can detect the distance enlargement of $P$. \label{fig:Minimum-distance-guarantee.}}

\end{figure}

Consequently, an abundance of protocols have been suggested to deal
with these various deficiencies. Song et al.~\cite{Song2008} give
an example of a secure distance bounding mechanism for VANETs, comprising
of three steps:
\begin{enumerate}
\item Traditional distance bounding is used to find the lower bound of the
distance between $V$ and $P$. $P$ can only increase the time to
respond to $V$'s challenge and as such only appear further away than
it really is.
\item The verifier then checks the claimed location of $P$ for plausibility
(also see Section \ref{sub:Traffic-Modeling-and-Plausibility-Checks}):

\begin{enumerate}
\item Transmission range-based verification: There are limits on the maximum
distance of wireless transmission in practice. Test runs can help
to find practical upper bounds for ADS-B users in a certain location/path.
If the prover claims to be further away, it can be considered malicious.
\item Speed-based verification: Considering the fact that the typical speed
of a given airplane in differing flight stages is known (certainly
the physically possible minimum and maximum velocity), consecutive
position claims have to be in a given window.
\end{enumerate}
\item To further improve the security, after all plausibility checks have
been passed, the verifier chooses a common neighbour B. That neighbour
of both $P$ and $V$ then gives its location estimation $E$ for
$P$ as shown in Fig.\,\ref{fig:Minimum-distance-guarantee.}. Whenever
the estimate lies outside the error margin, $B$ knows that $P$ enlarged
its distance.\emph{ }
\end{enumerate}
Chiang et al.~\cite{Chiang2009,Chiang2012} develop a secure multilateration
scheme based on distance bounding that, under idealized assumptions,
can detect false location claims with a high rate of success. By also
taking into account RSS differences, they can mitigate distance enlargement
attacks and generic collusion attacks on the protocol. This shows
how different physical layer techniques can be combined to improve
theoretical security, however, the practical challenges in using such
protocols in ATC are difficult to solve.

While in the literature distance bounding has been used mostly for
close-up, indoor communication, it has been modeled for use in VANETs
up to a distance of 225\,m between prover and verifier \cite{Song2008,Ranganathan}.
Tippenhauer and Capkun \cite{Tippenhauer2009} also considered the
impact of moving targets on distance bounding protocols and verifiable
multilateration. In their original implementation, it takes about
600ms to perform a full localization, which, at a speed of only 500\,km/h
means that a target already moves 75\,m during the process\emph{.
}The authors propose Kalman filters (see Section \ref{sub:Kalman-Filtering})
to smoothly keep track of the prover's location and detect any malicious
tampering by outsiders.

Besides its current unsuitability for the long distances and high
velocities present in air-traffic control, another main disadvantage
of distance bounding is the fact that it inherently requires a response
by the prover to the verifier's challenge and thus from an ADS-B point
of view enforces an entirely new protocol paradigm. As an additional,
on-demand feature it could still provide crucial information about
the legitimacy of nodes in areas where PSR is not present (or is phased
out due to cost reasons).\\

\subsection{Kalman Filtering and Intent Verification\label{sub:Kalman-Filtering}}

Kalman filters (also known under the technical term linear quadratic
estimation) \cite{Kalman1960a} have already seen extensive use in
broader ATC applications, e.g. to filter and smoothen GPS position
data in messages. Kalman filtering is used to observe noisy time series
of measurements and tries to statistically optimally predict future
states of the measured variables of the underlying system.\emph{ }

\begin{figure}
\begin{centering}
\includegraphics[scale=0.6]{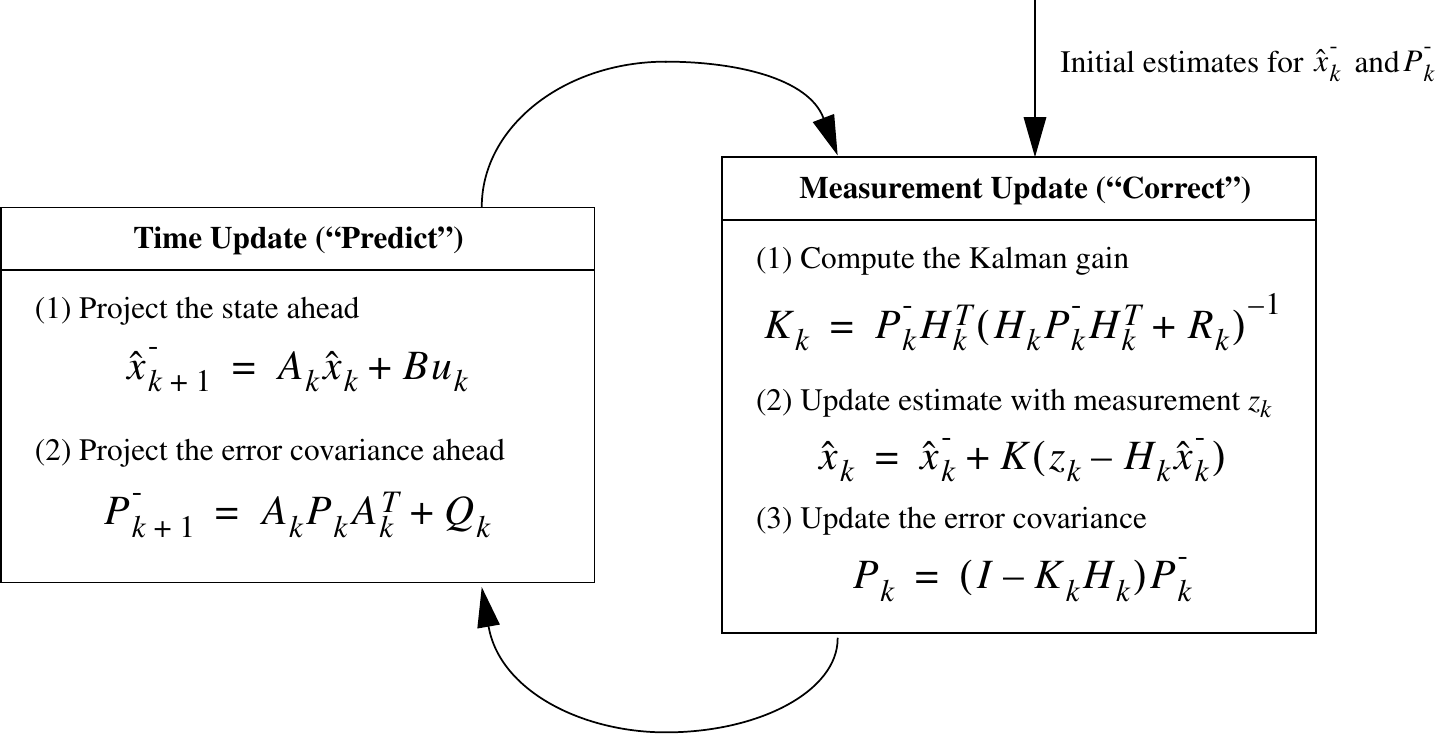}
\par\end{centering}

\caption{Basic concept of Kalman filtering from \cite{Welch1995}. The time
update step projects the measured variables and the error covariances.
The measurement update step computes the Kalman gain and updates the
estimates and error covariances with the actual measurements. \label{fig:Basic-concept-ofKalman}}
\end{figure}

A high level overview of the Kalman filtering algorithm comprises
two distinct steps, a prediction step and an update step (see Fig.\,\ref{fig:Basic-concept-ofKalman}).
As the procedure is recursive it can easily be used and updated in
real time, without having to save more than the last state. 
\begin{itemize}
\item Prediction step: In the first step, the current state variables are
predicted as well as the connected uncertainties.\emph{ }
\item Update step:\emph{ }For every following step,\emph{ }the previously
obtained estimates are then updated with a weighted average. During
this process, the estimates with higher certainty are assigned higher
weight.\emph{ }
\end{itemize}

The theory behind Kalman filtering requires the observed system to
be linear and the underlying measuring variables and errors to follow
a normal distribution, although there have been developments to adapt
the approach to non-linear systems.\emph{ }

\begin{figure}
\includegraphics[scale=0.27]{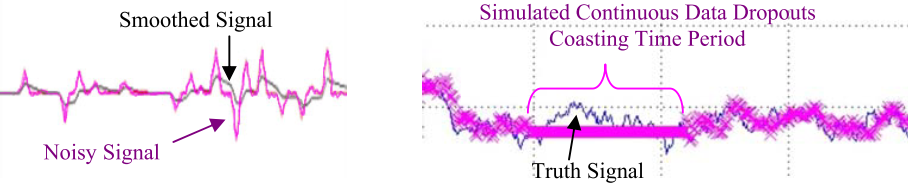}

\caption{Example of Kalman filtering from \cite{Krozel2004}. Noisy signals
are being smoothed (left), dropped data being coasted over below a
given cutoff point (right). \label{fig:Example-of-Kalman}}

\end{figure}

Kalman filtering plays a crucial role in the multilateration approach,
sorting out noisy signals and smoothing over missing data (Fig.\,\ref{fig:Example-of-Kalman}).
It is also a useful tool in general to predict the future values of
a feature based on collected historical data.\emph{ }More concretely,
it is used in ground systems to filter and verify the state vectors
and trajectory changes reported by ADS-B aircraft and conduct plausibility
checks on these data \cite{Fox2003}\emph{.} Krozel et al.~\cite{Krozel2004}
go on further to verify the intent of the aircraft by first defining
local and global correlation functions to evaluate the correlation
between aircraft motions and the ADS-B intent (Fig.\,\ref{fig:Basic-concept-ofKalman}).
Then the authors compute geometric conformance, i.e. if the aircraft
is in given horizontal and vertical limits and intent conformance,
i.e. analyzing the aircraft motion and comparing it to a plausible
intent model in several dimensions (in this case horizontal, vertical
and velocity).

\begin{figure}
\begin{centering}
\includegraphics[bb=10bp 5bp 610bp 125bp,clip,scale=0.42]{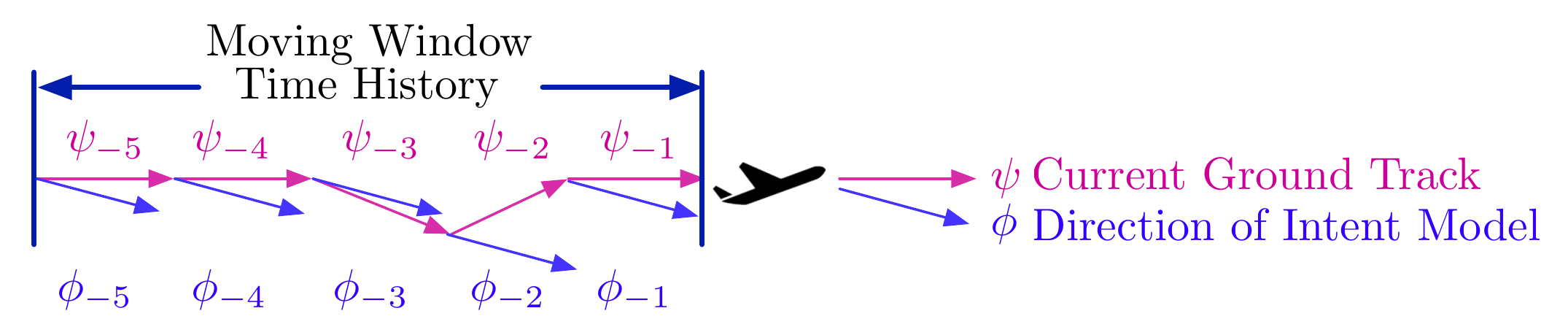}
\par\end{centering}

\caption{Practical application of intent verification in ADS-B from \cite{Krozel2004}.
The example analyzes the horizontal aircraft motion with a global
correlation function as a moving window over local correlation functions.}
\end{figure}

Kovell et al.~\cite{Kovell2012} note that since Kalman filtering
is used in a number of ADS-B related systems, it is essential to distinguish
between Kalman filters dealing with an aircraft's GPS position, with
received signal strength of packets and the angle of arrival at a
recipient's antenna and their proposed use for real time positional
claim verifications of an aircraft onboard of other aircraft. Kalman
filtering of positional claims is slightly more difficult in aircraft-to-aircraft
systems but there are no inherent impossible obstacles to it. 

From an attacker's point of view, Kalman filters can be tricked by
a so-called frog boiling attack \cite{Chan-tin2010}: The adversary
is jamming the correct signal, while continuously transmitting an
ever-so-slightly modified position. If this is done slowly enough,
the Kalman filter will see the injected data as a valid trajectory
change. This exposes a general weakness of Kalman filtering as the
approach is based on comparatively little historical data. But it
is still of great use since obviously bogus manoeuvres, speeds, features
can be detected (see also Section \ref{sub:Traffic-Modeling-and-Plausibility-Checks})
and the complexity of any attack is greatly increased. Another general
downside is that it opens up more DoS-possibilities due to the largely
increased computational complexity at every receiver, although this
is not a major problem with comparably powerful installations in ground
stations and airplanes. A possible threshold time after which sufficient
trust has been established between two participants based Kalman filtering
is still open research \cite{Kovell2012}. \\

\subsection{Group Verification}

Group verification is another concept proposed to mitigate security
and privacy concerns over the use of ADS-B \cite{Sampigethaya2011}.
It aims at securing the airborne ADS-B IN communication by employing
multilateration done by a group to verify location claims of non-group
members in-flight. A given authenticated group with 4 or more aircraft
having established trust can communicate with each other to utilize
multilateration (based on TDOA or RSS) just as ground stations can
(see Section \ref{sub:Multilateration}). If a forged position report
is detected, the sensible reaction would be to increase the circle
of avoidance around nearby airplanes since their position cannot anymore
be regarded as precisely known and thus safe.

Kovell et al.~\cite{Kovell2012} conducted a study about the applicability
of the group concept in commercial aviation in the United States airspace.
Examining the vast differences in traffic density over the US, they
found that around 91\% of aircraft at a given time could be part of
a sufficiently large group of 4 aircraft or more.

Group verification has a number of downsides. First of all, it requires
many additional messages to implement the verification and trust process.
As ADS-B is purely unidirectional broadcast, a new protocol is needed
to support the group concept. Concerning the question of which protocol
to use, the authors mention the L-Band Digital Aeronautical Communication
System (L-DACS) as one possibility. L-DACS is being developed by EUROCONTROL
as a future IP technology for air-to-air communication but unfortunately
there is no specification in sight in the medium term.%
\footnote{``In addition to the air/ground capability, some of the assessed
technologies could also support additional features such as air/air
(point to point and/or broadcast) communications and digital voice.
However the support of these capabilities needs further investigation.''
http://www.eurocontrol.int/communications/public/standard\_page/LDACS.html%
} If such a protocol can be successfully implemented, there remains
the central problem of how to manage the secure authentication of
members that are to be accepted into the group in the first place.
It is very complicated to establish trust in new groups of MANETs
and to reliably avoid malicious aircraft. Furthermore, the performance
of the system in reaction to intelligent intentional jamming of some
or all communication would have to be considered.

\begin{figure}
\includegraphics[scale=0.3]{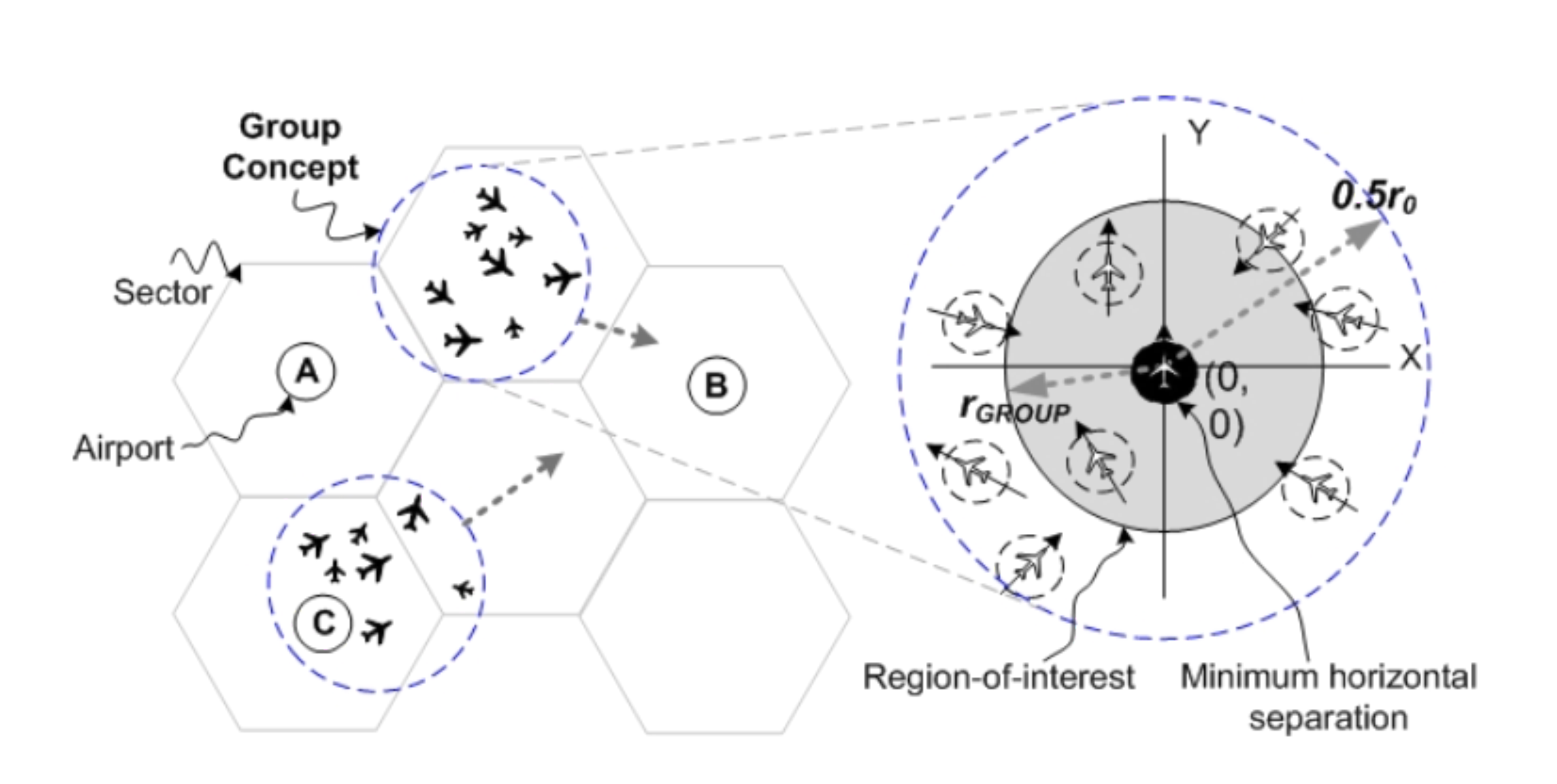}

\caption{Illustration of the group concept from \cite{Sampigethaya2011}. Four
or more aircraft ${V}$ are in any group $G$. Each group then can
internally use multilateration to verify each other's location claims
as well as those of outsiders in range. $r_{0}$ is the wireless communication
range, $0.5r_{0}$ considered geographically proximate and thus acceptable
for group establishment, given sufficient communication quality. To
lower group overhead, the region of interest for a group can be restricted
to $r_{group}$.}
\end{figure}

On the other hand, even without a perfectly secure solution, the group
concept would raise the difficulty and engineering effort of certain
airborne attacks by orders of magnitude. \\

\subsection{Data Fusion and Trust Management}

Data fusion is quickly becoming a cornerstone of modern intelligent
transport systems (ITS). The concept can be used at various stages
of data processing, Baud et al.~\cite{Baud2006}, for example, describe
the fusion of radar and ADS-B data and show that this approach can
improve the quality of tracking in practice. Concerning ADS-B security,
the literature proposes to check positional data obtained from within
the system against data coming in from other, independent sources.
Adequate data can e.g. stem from multilateration (see Section \ref{sub:Multilateration}),
traditional primary radar systems or even flight plan data. Liu et
al.~\cite{liu2013multi} describe a hybrid estimation algorithm to
fuse multiple sensors with different surveillance techniques (PSR,
SSR, multilateration) and flight plan information together, not specifically
for security purposes but general fault detection.

Such verification can provide a way of knowing if some of the involved
systems work outside normal parameters, be it from a malicious source
or not. Subsequently, automated technical or non-technical procedures
can be carried out, identifying the problem and reacting accordingly.
This process comprises an analysis of the trust-worthiness of the
data, if it has been vulnerable to tampering depending on the system
and the precision/measurement uncertainty of the respective technologies
(as given e.g. in \cite{Smith2006}). The trust-worthiness can then
be calculated by looking at the correlations and further features
deduced through machine learning processes which aim to expose anomalies
in received information and thus to enable more automated detection
of attacks.

An example of this is given in \cite{Wei2003}. The authors use the
cosine similarity between claimed and estimated positions to judge
the trustworthiness of a participant's claims and maintain historical
beacon trust information:

{\scriptsize 
\begin{equation}
Sim_{Cos}(\vec{E},\vec{O})=\frac{\vec{E}\cdot\vec{O}}{|\vec{E}|\cdot|\vec{O}|}=\frac{x_{E}\times x_{O}+y_{E}\times+v_{E}\times v_{O}}{\sqrt{x_{E}^{2}+y_{O}^{2}+v_{E}^{2}}\times\sqrt{x_{O}^{2}+y_{O}^{2}+v_{O}^{2}}}
\end{equation}
}{\scriptsize \par}

where $x_{O}$, $y_{O}$ are the coordinates, $v_{O}$ the velocity
as claimed in the last received message and $x_{E}$, $y_{E}$ are
the estimated coordinates of the claimant, based on the previously
received message. As a further step they calculate the time-based
weighted trustworthiness of a beacon message, taking into account
the cosine similarity of the last $I$ beacon messages and their respective
estimates:

{\scriptsize 
\begin{equation}
T_{beacon}=\frac{\sum_{i=1}^{I}Sim_{Cos}(\vec{E},\vec{O})(w_{i})^{n}}{\sum_{i=1}^{I}(w_{i})^{n}}
\end{equation}
}{\scriptsize \par}

It is easy to make a case for data fusion, since many of the required
components are already available and a two-out-of-three (2oo3) approach
is a widely accepted best practice in many industries dealing with
processes that are crucial to safety and security. It has even been
suggested to equip the average car with primary surveillance radar,
to secure vehicular ad hoc networks \cite{Yan2008}.

There are already a number of integrated systems being deployed (such
as ASDE-X) that are inherently fusing the data of various sub-systems
(ADS-B, multilateration, flight plan information, radar data) to increase
both security and accuracy in airport proximity. An exemplary patent
for a data fusion apparatus looking to improve ADS-B security can
be found in \cite{smith2008method}. The clear advantage of data fusion
is the compatibility with legacy systems, including the fact that
the ADS-B protocol does not need to be amended to provide these additional
features. The downsides include increased cost for additional systems
to provide the necessary redundancy.\\

\subsection{Traffic Modeling and Plausibility Checks \label{sub:Traffic-Modeling-and-Plausibility-Checks}}

\begin{figure}
\includegraphics[scale=0.3]{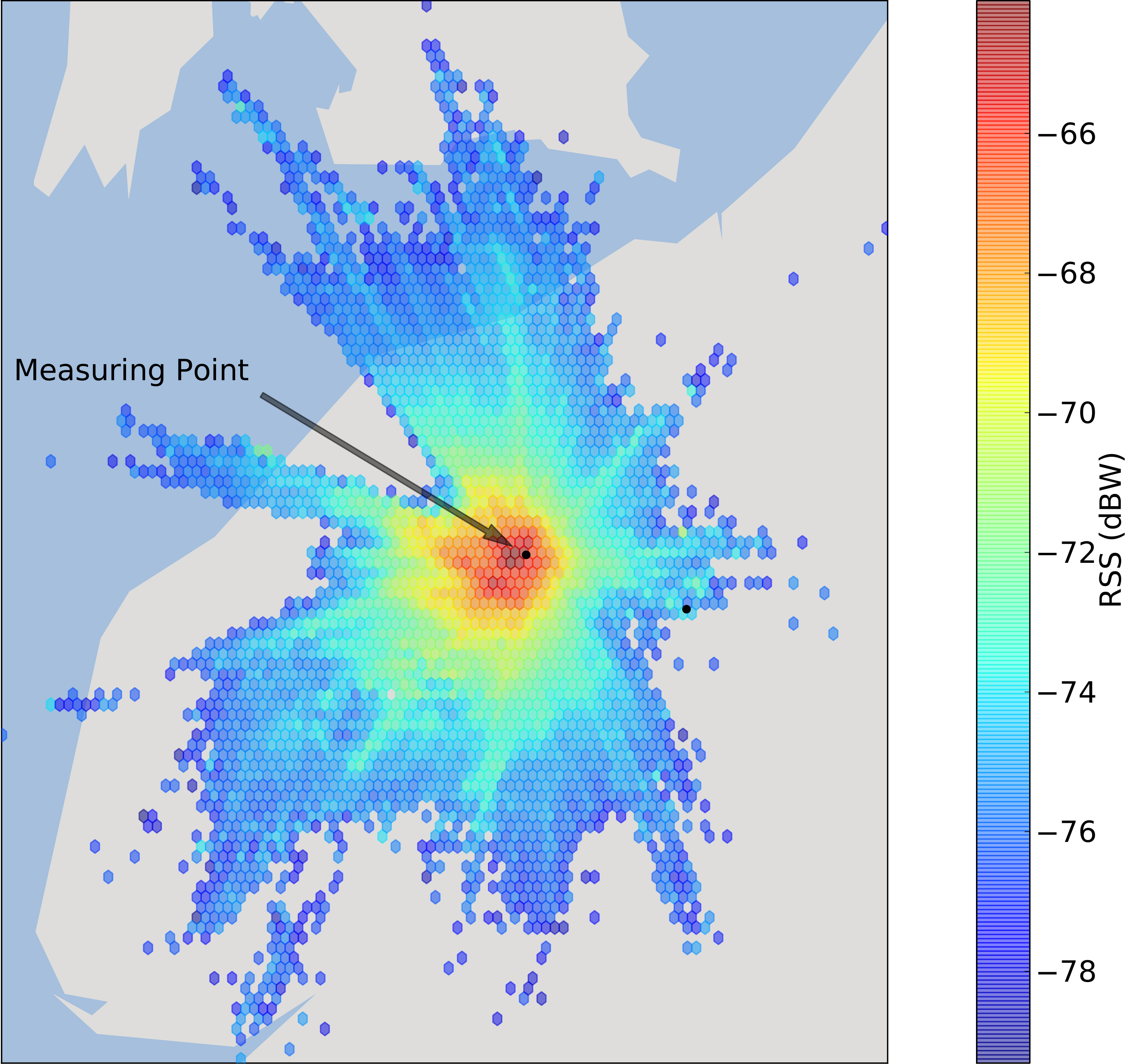}

\caption{Example of an RSS heat map used for traffic modeling from \cite{schafer2013experimental}.
The colors indicate the received signal strength at a single measuring
point which correlates strongly with the distance. Though it is relatively
easy for an attacker to manipulate the RSS, it becomes increasingly
difficult the more measuring points have to be deceived. \label{fig:Example-heat-map}}

\end{figure}

Traffic modeling could provide a mechanism to detect deviations from
normal ADS-B behaviour. By utilizing historical data as well as machine
learning methods it is possible to create a model of a map for each
ground station, providing a means to verify location claims made by
aircraft via ADS-B. Figure \ref{fig:Example-heat-map} shows how a
typical heat map based on RSS values looks from the point of view
of a ground station. Such models can provide hints about non-matching
location claims of a message send by an aircraft. Other potential
considerations include checking for a certain number of consecutive
packets of the same - or different - aircraft with the same RSS/angle-of-arrival,
or otherwise suspicious absolute values which could indicate a stationary,
ground-based attacker (e.g. RSS/AoA values outside typically observed
thresholds).

Xiao et al.~\cite{Xiao} used a similar statistical approach for
the detection of Sybil nodes in VANETs. They propose an algorithm
that could be performed by any node that has received enough measurements
of e.g. signal strength from nearby witnesses. While a single estimated
position of a node may not be an accurate representation of the real
location, a larger sample of estimated positions would have to be
very similar to the node's position claims over a given period of
time. 

More formally, if in a period $\Delta t_{o}$ there are $n$ sequential
positions $l_{1},...,l_{n}$ claimed by an airplane with the corresponding
estimated positions $l_{1}^{'},...,l_{n}^{'}$, then the difference
between the two can be treated as a random error. That means that
a large enough sample of differences is distributed normally with
a mean $\mu_{d}$ and a variance $\sigma_{0}^{2}$. If these hypotheses
are true, then it can be assumed that the claimant has given valid
positional reports within the chosen level of significance (see \cite{Xiao}
for further explanations and simulation results in VANETs). As this
method is only practical for fixed receivers, the goal is to use ground
stations to detect unusual behaviour, which are also more likely to
be able to collect enough samples to make the verification process
as sound as possible.

On top of this, there are numerous comparably simple rules that can
be utilized as potential red flags by an intrusion detection system
without resorting to more complex measures. Neither of these rules
are necessary nor sufficient by themselves to detect an ongoing attack.
But depending on the scenario and the attacker's savvy, they can indicate
unusual behaviour that should be further investigated either by a
human or handled through additional technical means. For example,
it is very plausible to outright drop a number of packets where either
the data or the meta data is technically or physically impossible.
Doing so can significantly reduce the strain on the ADS-B system and
even prevent both spoofing and DoS attacks which are not crafted carefully
enough. If a large number of potential red flags are checked for an
intrusion detection system, not only the risk for an attacker to cause
an alert increases significantly, but also the cost and complexity
of an attack rise. While this does not constitute theoretical perfect
security, plausibility checks can be very useful in practice. 

Mitigating factors exist across various layers, from the physical
to the application. Some are available to ground stations/air traffic
control only, others also to aircraft-to-aircraft (A2A) ADS-B IN communication.
Such cases include but are not limited to \cite{Leinmuller2006}:\\

\begin{itemize}
\item Investigating\textbf{ airplanes which suddenly appear} well within
the maximum communication range of a receiver.
\item Dropping \textbf{aircraft which are violating a given acceptance range
threshold}, producing impossible locations.
\item \textbf{Aircraft violating a given mobility grade threshold}, producing
impossible minimum or maximum velocities.
\item \textbf{Maximum Density Threshold}: If too many aircraft are in a
given area, ATC software will typically alarm the user.
\item \textbf{Map-based Verification}: Aircraft in unusual places such as
no-fly areas or outside typical airways (this might possibly be better
handled at the ATC software layer).
\item \textbf{Flight plan-based Verification}: Flooded/attacked ground stations
are able to check ADS-B messages against the existing flight plan.
\item Obvious \textbf{discontinuities in one of the 9 ADS-B state vector
data fields} (also see the related Section \ref{sub:Kalman-Filtering})
.
\end{itemize}
As explained before, such potential red flags need to be handled with
utmost care and typically they can not be automated but require much
additional scrutiny before any action is taken (e.g., the packet/flight
is considered an attack and dropped from ATC monitors). Yet, they
also enable the opportunity to follow up and activate further means
to secure the airspace. For example, when using such centralized detection
at the ground stations, these same stations could then destroy messages
sent by the detected offenders as outlined in \cite{Martinovic2009,Wilhelm2011}.\\

\section{Summary \label{sec:Summary}}

The Tables \ref{tab:Overview1}-\ref{tab:Overview3} provide a compact
overview over the effectiveness of all examined solutions in combating
the various proposed attacks as well as in offering advantages and
disadvantages concerning the feasibility to implement each approach
in the real world. As it has been laid out, there is no single optimal
or even good solution when considering means that have no or little
impact on the currently employed ADS-B software and hardware. Table
\ref{tab:Overview1} shows the attacks the discussed approaches can
counteract. We see that most security schemes focus on attacks of
the message injection/modification class. This has two main reasons
that have been mentioned throughout this survey: First of all, the
open nature of ADS-B has been considered a desirable feature in most
scenarios. So unless there is a major paradigm shift in the way air
traffic communication and control is handled currently, there is no
interest in protecting against passive listeners, despite this being
the first stepping stone for more sophisticated and problematic attacks.
Second, passive attacks such as eavesdropping are simply much more
difficult to protect against without having a full cryptographic solution.
Similarly, attacks on the physical layer, such as continuously jamming
the well-known frequency or the more surgical message deletion are
hard to defend against, with measures on the same layer (e.g. uncoordinated
spread spectrum) providing some of the only approaches to this general
wireless security problem. All discussed approaches do however address
message insertion and tampering, either by protecting outright against
it through verification (cryptographic methods) or by detecting anomalies
in the data (e.g.~Kalman filtering, multilateration).
\begin{table}[t]
\begin{tabular}{|>{\centering}p{24mm}|>{\centering}p{1.5cm}|c|>{\centering}p{1.5cm}|}
\hline 
 & Injection / Modification & Eavesdropping & Jamming / Deletion\tabularnewline
\hline 
\hline 
Physical Layer Authentication & + & - & -\tabularnewline
\hline 
Uncoordinated Spread Spectrum & - & + & +\tabularnewline
\hline 
(Lightweight) PKI & + & + & -\tabularnewline
\hline 
μTESLA & + & - & -\tabularnewline
\hline 
Wide Area Multilateration & + & - & -\tabularnewline
\hline 
Distance Bounding & + & - & -\tabularnewline
\hline 
Kalman Filtering & + & - & -\tabularnewline
\hline 
Group Verification & + & - & -\tabularnewline
\hline 
Data Fusion & + & - & -\tabularnewline
\hline 
Traffic Modeling & + & - & -\tabularnewline
\hline 
\end{tabular}

\caption{Overview of capabilities of various security approaches against feasible
attacks on ADS-B. \label{tab:Overview1}}
\end{table}
\begin{table}[t]
\begin{tabular}{|>{\centering}p{2.4cm}|>{\centering}p{1.1cm}|>{\centering}p{1.1cm}|>{\centering}p{1.1cm}|>{\centering}p{1.1cm}|}
\hline 
 & Data Integrity & Source Integrity & Location Integrity & DoS\tabularnewline
\hline 
\hline 
Physical Layer Authentication & No & Yes & Possibly & Partly\tabularnewline
\hline 
Uncoordinated Spread Spectrum & No & No & No & Yes\tabularnewline
\hline 
(Lightweight) PKI & Yes & Yes & Yes & Partly\tabularnewline
\hline 
μTESLA & No & Yes & No & No\tabularnewline
\hline 
Wide Area Multilateration & No & No & Yes & No\tabularnewline
\hline 
Distance Bounding & No & No & Partly & No\tabularnewline
\hline 
Kalman Filtering & No & Partly & Partly & No\tabularnewline
\hline 
Group Verification & No & Possibly & Yes & No\tabularnewline
\hline 
Data Fusion & No & Partly & Yes & Backup\tabularnewline
\hline 
Traffic Modeling & No & No & Yes & No\tabularnewline
\hline 
\end{tabular}

\caption{Overview of security features of various approaches for use with ADS-B.
\label{tab:Overview2}}
\end{table}

Table \ref{tab:Overview2} takes a look at the security features the
discussed schemes can provide. As discussed before, only a full cryptographic
public key infrastructure can guarantee the integrity of received
data. All other approaches either aim to secure the integrity of the
source (e.g. μTESLA, many of the discussed physical layer schemes)
or seek to verify the provided location data independently. Additional
protection against flood denial of service attacks against ATC systems
can be directly provided by spread spectrum approaches and cryptography,
while other methods rely on higher layers to sort out false aircraft
claims.

Table \ref{tab:Overview3} provides an overview over the feasibility
of the different approaches in practical settings, especially considering
the current state of air traffic control in the aviation industry.
As is to be expected, the difficulty and cost columns are mostly correlated.
The difficulty to overcome technical challenges is particularly high
for distance bounding, which is yet in its beginnings and a full-blown
public key infrastructure. In contrast, we see wide-area multilateration,
Kalman filters and data fusion techniques already in use in the field.
This naturally translates to the cost factor which plays an important
role in industry decisions. One can choose between a completely new
protocol which addresses the security question better than the current
installment of ADS-B does, slight modifications such as new message
types, or a transparent, parallel system which requires new software
and/or new hardware in different scales. This touches also on the
question of scalability. For example, the fusion of various ATC systems
and their data (PSR, SSR, ADS-C, WAMLAT, FANS) is an obvious and necessary
idea. Yet, it is common knowledge in the aviation community that a
``major part of the business case for Automatic Dependent Surveillance
- Broadcast (ADS-B) is attributed to the savings generated by decommissioning
or reducing reliance on conventional radar systems`` \cite{Smith2006}.
Thus, it seems inefficient and unlikely to have legacy and/or new
backup systems around on a broad scale, simply to fix the inadequateness
of ADS-B related to security.
\begin{table*}[t]
\begin{centering}
\begin{tabular}{|>{\centering}p{2.5cm}|>{\centering}p{1.3cm}|>{\centering}p{1.2cm}|>{\centering}p{1.2cm}|>{\centering}p{4cm}|>{\centering}p{3cm}|}
\hline 
 & Difficulty  & Cost & Scalability & Compatibility & References\tabularnewline
\hline 
\hline 
Physical Layer 

Authentication & Variable & Variable & Variable & Requires additional hard-/software. No modifications to the ADS-B
protocol. & \cite{Danev2012,Danev2010,Zeng2010,hall2005radio,Jana2010,Maes2010,Devadas2008,Mathur2008,Zhang2010a,Wang2011a,Xiong,Qiu,Laurendeau2008,Laurendeau2009}\tabularnewline
\hline 
Uncoordinated Spread Spectrum & Medium & Medium & Medium & Requires new hardware and a new physical layer. & \cite{Strasser2008,Christina2009,Liu2010}\tabularnewline
\hline 
(Lightweight) PKI & High & High & Medium & Distribution infrastructure and changes in protocol and message handling
needed.  & \cite{Costin,Finke2013a,wessoncan,Samuelson2006,viggiano2010secure,schuchman2011automatic,Ziliang2010,Raya2005,Raya2007,Robinson2007,Parno,Zhang2010}\tabularnewline
\hline 
μTESLA & Medium & Medium & High & New message type required for key publishing, MAC added. & \cite{Perrig2003,Perrig2000,Perrig2005,Perrig2002,Liu,Eldefrawy2010,Haas2009,Hu2006}\tabularnewline
\hline 
Wide Area Multi\-lateration & Low & Medium & Medium & No change to ADS-B required. Separate hardware system. & \cite{Smith2006,Savvides2002,Neven2005,Purton2010,Johnson2012,Kaune2012,Thomas2011a,Daskalakis2003,Niles2012,Galati2005}\tabularnewline
\hline 
Distance Bounding & High & Medium & Low & New messages and protocol needed. & \cite{Brands1994,Song2008,Chiang2009,Chiang2012,Ranganathan,Tippenhauer2009}\tabularnewline
\hline 
Kalman Filtering & Low & Low & High & No additional messages needed. Separate software system. & \cite{Kovell2012,Kalman1960a,Welch1995,Fox2003,Krozel2004}\tabularnewline
\hline 
Group Verification & High & Medium & Low & New messages and protocol needed. & \cite{Sampigethaya2011,Kovell2012}\tabularnewline
\hline 
Data Fusion & Low & Medium - High & Medium & No change in ADS-B required. Separate system. & \cite{Baud2006,liu2013multi,Wei2003,Yan2008,smith2008method}\tabularnewline
\hline 
Traffic Modeling & Medium & Low & High & Additional, separate entities for ground stations needed. & \cite{schafer2013experimental,Xiao,Leinmuller2006}\tabularnewline
\hline 
\end{tabular}
\par\end{centering}

\caption{Overview of feasibility attributes of various approaches for use with
ADS-B. \label{tab:Overview3}}
\end{table*}

\subsection*{Future Research Directions}

Considering the fact that the invention, certification and large-scale
deployment of air-traffic systems takes decades, as currently seen
in the example of ADS-B, it seems equally non-sensible to present
a completely overhauled ATC-system at this point in time. Yet, future
aviation protocol development is important as there will eventually
be a successor to ADS-B and the responsible community must learn from
the ADS-B case study. Authentication should be considered right from
the beginning when planning new protocols, this includes choosing
the right cryptographic primitives, an appropriate communication pattern
considering A2A and ground stations, and, most importantly, a solution
to the key management problem. Furthermore, the challenges and realities
of communication in the avionic environment need to be taken into
account, for example the extremely lossy environment (as recently
examined and illustrated with OpenSky, an open research sensor network
in \cite{Strohmeier14,Schaefer14}) that might render many traditional
approaches infeasible.

For the urgent problem at hand, this means that incremental changes
with backwards compatibility kept in mind as a main factor are more
useful than completely new proposals. Improvement on transparent secure
location verification approaches such as multilateration can help
bridge the security gap in the very near future, although this means
losing some of the advantages for which ADS-B was originally developed.
In the same vain, fingerprinting methods on various layers can help
build an effective intrusion detection system against all but the
most sophisticated attackers without affecting the protocol as it
is currently deployed. Improvements on current data fusion algorithms
can also further both safety and security of currently deployed ATC
by reducing error margins and uncertainties for controllers.

Of course, cost and complexity of deployment cannot be the only factors
taken into consideration - not having security is famously even much
more expensive.%
\footnote{``If you think safety is expensive, try an accident”, an insight
by easyJet owner Stelios Haji-Ioannou that came after facing manslaughter
charges for the deaths of employees in a shipping disaster at a former
company.%
} As mentioned throughout this survey, it can be a useful dichotomy
to distinguish between attack detection, attack prevention and dealing
with suspected attacks. We focused mainly on attack detection and
prevention, leaving attack reaction for future research. \\

\section{Conclusion \label{sec:Conclusion}}

This survey sought to review the available research on the topic of
securing the ADS-B protocol in particular and air traffic control
communication in general. We provided an in-depth overview of the
existing work, both specific to ADS-B as well as ideas brought in
from related fields such as VANETs. After reviewing the literature,
it seems that the solutions currently under consideration (and in
use in practice such as multilateration) can only be a fill-in, providing
a quick improvement to the security of the current system. For all-encompassing
security (and possibly privacy), new message types and/or completely
new protocols would need to be defined. Taking this into account for
the creation of a long-term security solution in dependent air traffic
surveillance, it makes sense to consider the impact of both secure
broadcast authentication approaches as well as of secure location
verification. To avoid new hard challenges in the foreseeable future,
this should include a thorough analysis of the predicted traffic density
on today's wireless navigation channels as well as the possible impact
of the communication and message overhead of a new protocol.\\

\bibliographystyle{IEEEtran}
\bibliography{ads-b_overview_revision}

% Generated by IEEEtran.bst, version: 1.12 (2007/01/11)
\begin{thebibliography}{100}
\providecommand{\url}[1]{#1}
\csname url@samestyle\endcsname
\providecommand{\newblock}{\relax}
\providecommand{\bibinfo}[2]{#2}
\providecommand{\BIBentrySTDinterwordspacing}{\spaceskip=0pt\relax}
\providecommand{\BIBentryALTinterwordstretchfactor}{4}
\providecommand{\BIBentryALTinterwordspacing}{\spaceskip=\fontdimen2\font plus
\BIBentryALTinterwordstretchfactor\fontdimen3\font minus
  \fontdimen4\font\relax}
\providecommand{\BIBforeignlanguage}[2]{{%
\expandafter\ifx\csname l@#1\endcsname\relax
\typeout{** WARNING: IEEEtran.bst: No hyphenation pattern has been}%
\typeout{** loaded for the language `#1'. Using the pattern for}%
\typeout{** the default language instead.}%
\else
\language=\csname l@#1\endcsname
\fi
#2}}
\providecommand{\BIBdecl}{\relax}
\BIBdecl

\bibitem{NewScientist}
\BIBentryALTinterwordspacing
P.~Marks. (2011, Sep) Air traffic system vulnerable to cyber attack. New
  Scientist. [Online]. Available:
  \url{http://www.newscientist.com/article/mg21128295.600-air-traffic-system-vulnerable-to-cyber-attack.html}
\BIBentrySTDinterwordspacing

\bibitem{CNN}
\BIBentryALTinterwordspacing
H.~Kelly. (2012, Jul.) Researcher: New air traffic control system is hackable.
  CNN. [Online]. Available:
  \url{http://edition.cnn.com/2012/07/26/tech/web/air-traffic-control-security/index.html}
\BIBentrySTDinterwordspacing

\bibitem{Forbes}
\BIBentryALTinterwordspacing
A.~Greenberg. (2012, Jul.) Next-gen air traffic control vulnerable to hackers
  spoofing planes out of thin air. Forbes. [Online]. Available:
  \url{http://www.forbes.com/sites/andygreenberg/2012/07/25/next-gen-air-traffic-control-vulnerable-to-hackers-spoofing-planes-out-of-thin-air/}
\BIBentrySTDinterwordspacing

\bibitem{Wired}
\BIBentryALTinterwordspacing
K.~Zetter. (2012, Jul.) Air traffic controllers pick the wrong week to quit
  using radar. Wired. [Online]. Available:
  \url{http://www.wired.com/threatlevel/2012/07/adsb-spoofing/}
\BIBentrySTDinterwordspacing

\bibitem{NPR}
\BIBentryALTinterwordspacing
S.~Henn. (2012, Aug.) Could the new air traffic control system be hacked? NPR.
  [Online]. Available:
  \url{http://www.npr.org/blogs/alltechconsidered/2012/08/16/158758161/could-the-new-air-traffic-control-system-be-hacked}
\BIBentrySTDinterwordspacing

\bibitem{Costin}
A.~Costin and A.~Francillon, ``{Ghost in the Air (Traffic): On insecurity of
  ADS-B protocol and practical attacks on ADS-B devices},'' in \emph{Black Hat
  USA}, 2012.

\bibitem{bradhaines2012}
B.~Haines, ``Hacker + airplanes = no good can come of this,'' in
  \emph{Confidence X}, 2012.

\bibitem{defconkunkel}
R.~Kunkel, ``Air traffic control insecurity 2.0,'' in \emph{DefCon 18}, 2010.

\bibitem{schafer2013experimental}
M.~Sch\"{a}fer, V.~Lenders, and I.~Martinovic, ``{Experimental Analysis of
  Attacks on Next Generation Air Traffic Communication},'' in \emph{Applied
  Cryptography and Network Security}.\hskip 1em plus 0.5em minus 0.4em\relax
  Springer, 2013, pp. 253--271.

\bibitem{ICAO2012}
ICAO, ``{Cyber Security for Civil Aviation},'' in \emph{Twelfth Air Navigation
  Conference}, 2012, pp. 1--4.

\bibitem{adsbimplementation}
------, ``{Status of ADS-B Avionics Equipage Along ATS Routes L642/M771 For
  Harmonized ADS-B Implementation},'' in \emph{ADS-B Seminar and 11th Meeting
  of ADS-B Study and Implementation Task Force}, Apr. 2012.

\bibitem{airbusstatus}
L.~Vidal, ``{ADS-B Out and In - Airbus Status},'' ADS-B Taskforce - KOLKATA,
  Apr. 2013.

\bibitem{cascade}
EUROCONTROL, ``{CASCADE News 9 - Update on Developments},'' Oct. 2010.

\bibitem{Smith2006}
A.~Smith, R.~Cassell, T.~Breen, R.~Hulstrom, and C.~Evers, ``{Methods to
  Provide System-wide ADS-B Back-Up, Validation and Security},'' in \emph{25th
  Digital Avionics Systems Conference}, 2006, pp. 1--7.

\bibitem{DO2422}
{RTCA Inc.}, ``{Minimum Aviation System Performance Standards for Automatic
  Dependent Surveillance Broadcast (ADS-B)},'' DO-242A (including Change 1),
  Dec. 2006.

\bibitem{DO260B}
------, ``{Minimum Operational Performance Standards for 1090 MHz Extended
  Squitter Automatic Dependent Surveillance -- Broadcast (ADS-B) and Traffic
  Information Services -- Broadcast (TIS-B)},'' DO-260B with Corrigendum 1,
  Dec. 2011.

\bibitem{DO282B2}
------, ``{Minimum Operational Performance Standards for Universal Access
  Transceiver (UAT) Automatic Dependent Surveillance -- Broadcast},'' DO-282B
  with Corrigendum 1, Dec. 2011.

\bibitem{McCallie2011}
D.~McCallie, J.~Butts, and R.~Mills, ``{Security analysis of the ADS-B
  implementation in the next generation air transportation system},''
  \emph{International Journal of Critical Infrastructure Protection}, vol.~4,
  no.~2, pp. 78--87, Aug. 2011.

\bibitem{barry}
J.~R. Barry, \emph{{Wireless Infrared Communications}}.\hskip 1em plus 0.5em
  minus 0.4em\relax Boston: Kluwer Academic, 1994.

\bibitem{Skolnik07}
{Merrill Ivan Skolnik}, \emph{{Radar handbook}}, 3rd~ed.\hskip 1em plus 0.5em
  minus 0.4em\relax {McGraw-Hill Professional}, 2007.

\bibitem{ICAO2007}
ICAO, ``{Guidance Material on Comparison of Surveillance Technologies
  (GMST)},'' Tech. Rep. September, 2007.

\bibitem{Strohmeier14}
M.~Strohmeier, M.~Sch{\"a}fer, V.~Lenders, and I.~Martinovic, ``{Realities and
  Challenges of NextGen Air Traffic Management: The Case of ADS-B},''
  \emph{Communications Magazine, IEEE}, vol.~52, no.~5, 2014.

\bibitem{Wilhelm2011a}
M.~Wilhelm and I.~Martinovic, ``{Short paper: reactive jamming in wireless
  networks: how realistic is the threat?}'' \emph{Proceedings of the fourth ACM
  conference on Wireless network security}, pp. 47--52, 2011.

\bibitem{Purton2010}
L.~Purton, H.~Abbass, and S.~Alam, ``{Identification of ADS-B System
  Vulnerabilities and Threats},'' in \emph{Australian Transport Research Forum,
  Canberra}, 2010, pp. 1--16.

\bibitem{Popper2011}
C.~P\"{o}pper, N.~O. Tippenhauer, B.~Danev, and S.~\v{C}apkun, ``{Investigation
  of signal and message manipulations on the wireless channel},'' in
  \emph{ESORICS'11 Proceedings of the 16th European Conference on Research in
  Computer Security}, 2011.

\bibitem{Wilhelm2012}
M.~Wilhelm, J.~B. Schmitt, and V.~Lenders, ``{Practical message manipulation
  attacks in IEEE 802.15.4 wireless networks},'' in \emph{MMB \& DFT 2012
  Workshop Proceedings}, 2012.

\bibitem{adamy2001ew}
D.~Adamy, \emph{EW 101: A first course in electronic warfare}.\hskip 1em plus
  0.5em minus 0.4em\relax Artech House, 2001.

\bibitem{Li2010}
H.~Li, B.~Yang, C.~Chen, and X.~Guan, ``{Connectivity of Aeronautical Ad hoc
  Networks},'' in \emph{2010 IEEE Globecom Workshops}.\hskip 1em plus 0.5em
  minus 0.4em\relax IEEE, Dec. 2010, pp. 1788--1792.

\bibitem{Boeing}
\BIBentryALTinterwordspacing
Boeing. (2013) Boeing current market outlook 2013-2032. [Online]. Available:
  \url{http://www.boeing.com/commercial/cmo/}
\BIBentrySTDinterwordspacing

\bibitem{Perrig2003}
A.~Perrig and D.~Tygar, \emph{{Secure Broadcast Communication in Wired and
  Wireless Networks}}.\hskip 1em plus 0.5em minus 0.4em\relax Springer, 2003.

\bibitem{haley2008security}
C.~B. Haley, R.~Laney, J.~D. Moffett, and B.~Nuseibeh, ``Security requirements
  engineering: A framework for representation and analysis,'' \emph{Software
  Engineering, IEEE Transactions on}, vol.~34, no.~1, pp. 133--153, 2008.

\bibitem{Martone2001a}
P.~J. Martone and G.~E. Tucker, ``{Candidate requirements for multilateration
  and ADS-B systems to serve as alternatives to secondary radar},'' in
  \emph{Digital Avionics Systems, 2001. DASC. 20th Conference}, 2001, pp.
  1--12.

\bibitem{Rekkas2008}
C.~Rekkas and M.~Rees, ``{Towards ADS-B implementation in Europe},'' in
  \emph{2008 Tyrrhenian International Workshop on Digital Communications -
  Enhanced Surveillance of Aircraft and Vehicles}.\hskip 1em plus 0.5em minus
  0.4em\relax IEEE, Sep. 2008, pp. 1--4.

\bibitem{Sampigethaya2011}
K.~Sampigethaya and R.~Poovendran, ``{Security and privacy of future aircraft
  wireless communications with offboard systems},'' in \emph{2011 Third
  International Conference on Communication Systems and Networks (COMSNETS
  2011)}.\hskip 1em plus 0.5em minus 0.4em\relax IEEE, Jan. 2011, pp. 1--6.

\bibitem{Sampigethaya}
K.~Sampigethaya and L.~Bushnell, ``{A Framework for Securing Future e-Enabled
  Aircraft Navigation and Surveillance},'' in \emph{AIAA Proceedings}, 2009,
  pp. 1--10.

\bibitem{Kovell2012}
B.~Kovell, B.~Mellish, T.~Newman, and O.~Kajopaiye, ``{Comparative Analysis of
  ADS-B Verification Techniques},'' 2012.

\bibitem{Nuseibeh2009}
B.~Nuseibeh, C.~B. Haley, and C.~Foster, ``{Securing the Skies: In Requirements
  We Trust},'' \emph{Computer}, vol.~42, no.~9, pp. 64--72, 2009.

\bibitem{Burbank2005}
J.~Burbank, R.~Nichols, S.~Munjal, R.~Pattay, and W.~Kasch, ``{Advanced
  communications networking concepts for the National Airspace System},'' in
  \emph{2005 IEEE Aerospace Conference}.\hskip 1em plus 0.5em minus 0.4em\relax
  IEEE, 2005, pp. 1905--1923.

\bibitem{Li2011}
W.~W. Li and P.~Kamal, ``{Integrated Aviation Security for Defense-in-Depth of
  Next Generation Air Transportation System},'' in \emph{IEEE Conference on
  Technologies for Homeland Security}, 2011, pp. 136--142.

\bibitem{Ochieng2003}
W.~Y. Ochieng, K.~Sauer, D.~Walsh, G.~Brodin, S.~Griffin, and M.~Denney, ``{GPS
  Integrity and Potential Impact on Aviation Safety},'' \emph{Journal of
  Navigation}, vol.~56, no.~1, pp. 51--65, Jan. 2003.

\bibitem{Siu2011}
J.~C. Siu, ``{ICAO Concepts and References Regarding ADS-B, Multilateration and
  Other Surveillance Techniques},'' in \emph{ICAO/FAA Workshop on ADS-B and
  Multilateration Implementation}, Sep. 2011.

\bibitem{Wagner2009}
R.~Wagner, ``{IFF, Combat ID and the Information Domino Effect},'' Canadian
  Department of National Defence, Tech. Rep., 2009.

\bibitem{Assessment1993a}
{U.S. Congress, Office of Technology Assessment}, \emph{{Who Goes There: Friend
  or Foe?}}\hskip 1em plus 0.5em minus 0.4em\relax Washington, DC: U.S.
  Government Printing Office, 1993.

\bibitem{Kenney2008}
L.~Kenney, J.~Dietrich, and J.~Woodall, ``{Secure ATC surveillance for military
  applications},'' in \emph{MILCOM 2008 - 2008 IEEE Military Communications
  Conference}.\hskip 1em plus 0.5em minus 0.4em\relax IEEE, Nov. 2008, pp.
  1--6.

\bibitem{DO209}
{RTCA Inc.}, ``{Proposed Change to DO-181D and ED-73C for Higher Squitter Rates
  at Lower Power},'' Special Committee 209 ATCRBS / Mode S Transponder MOPS
  Maintenance, Apr. 2007.

\bibitem{Luk2006}
M.~Luk, A.~Perrig, and B.~Whillock, ``{Seven cardinal properties of sensor
  network broadcast authentication},'' in \emph{Proceedings of the fourth ACM
  workshop on Security of ad hoc and sensor networks - SASN '06}.\hskip 1em
  plus 0.5em minus 0.4em\relax New York, New York, USA: ACM Press, 2006, p.
  147.

\bibitem{Woo1992}
T.~Y. Woo and S.~S. Lam, ``{Authentication for Distributed Systems},''
  \emph{Computer}, vol.~25, no.~1, pp. 39--52, 1992.

\bibitem{Danev2012}
B.~Danev, D.~Zanetti, and S.~Capkun, ``{On physical-layer identification of
  wireless devices},'' \emph{ACM Computing Surveys}, vol.~45, no.~1, pp. 1--29,
  Nov. 2012.

\bibitem{Danev2010}
B.~Danev, H.~Luecken, S.~Capkun, and K.~{El Defrawy}, ``{Attacks on
  physical-layer identification},'' in \emph{Proceedings of the third ACM
  conference on Wireless network security}, 2010, pp. 89--98.

\bibitem{Zeng2010}
K.~Zeng, K.~Govindan, and P.~Mohapatra, ``{Non-Cryptographic Authentication and
  Identification in Wireless Networks},'' \emph{IEEE Wireless Communications},
  pp. 1--8, 2010.

\bibitem{hall2005radio}
J.~Hall, M.~Barbeau, and E.~Kranakis, ``Radio frequency fingerprinting for
  intrusion detection in wireless networks,'' \emph{IEEE Transactions on
  Defendable and Secure Computing}, 2005.

\bibitem{Jana2010}
S.~Jana and S.~K. Kasera, ``{On Fast and Accurate Detection of Unauthorized
  Wireless Access Points Using Clock Skews},'' \emph{IEEE Transactions on
  Mobile Computing}, vol.~9, no.~3, pp. 449--462, Mar. 2010.

\bibitem{Maes2010}
R.~Maes and I.~Verbauwhede, ``Physically unclonable functions: A study on the
  state of the art and future research directions,'' in \emph{Towards
  Hardware-Intrinsic Security}.\hskip 1em plus 0.5em minus 0.4em\relax
  Springer, 2010, pp. 3--37.

\bibitem{Devadas2008}
S.~Devadas, E.~Suh, S.~Paral, R.~Sowell, T.~Ziola, and V.~Khandelwal, ``{Design
  and Implementation of PUF-Based "Unclonable" RFID ICs for Anti-Counterfeiting
  and Security Applications},'' in \emph{2008 IEEE International Conference on
  RFID}.\hskip 1em plus 0.5em minus 0.4em\relax IEEE, Apr. 2008, pp. 58--64.

\bibitem{Mathur2008}
S.~Mathur, W.~Trappe, N.~Mandayam, C.~Ye, and A.~Reznik, ``{Radio-telepathy:
  Extracting a Secret Key from an Unauthenticated Wireless Channel},'' in
  \emph{Proceedings of the 14th ACM international conference on Mobile
  computing and networking}, 2008, pp. 128----139.

\bibitem{Zhang2010a}
J.~Zhang, S.~K. Kasera, and N.~Patwari, ``{Mobility Assisted Secret Key
  Generation Using Wireless Link Signatures},'' in \emph{2010 Proceedings IEEE
  INFOCOM}.\hskip 1em plus 0.5em minus 0.4em\relax IEEE, Mar. 2010, pp. 1--5.

\bibitem{Wang2011a}
Q.~Wang, H.~Su, K.~Ren, and K.~Kim, ``{Fast and scalable secret key generation
  exploiting channel phase randomness in wireless networks},'' in \emph{2011
  Proceedings IEEE INFOCOM}.\hskip 1em plus 0.5em minus 0.4em\relax IEEE, Apr.
  2011, pp. 1422--1430.

\bibitem{Xiong}
J.~Xiong and K.~Jamieson, ``{SecureAngle: Improving Wireless Security Using
  Angle-of-Arrival Information},'' in \emph{Proceedings of the Ninth ACM
  SIGCOMM Workshop on Hot Topics in Networks}, 2010, p.~11.

\bibitem{Qiu}
D.~Qiu, D.~D. Lorenzo, S.~Lo, D.~Boneh, and P.~Enge, ``{Physical Pseudo Random
  Function in Radio Frequency Sources for Security},'' in \emph{Proceeding of
  ION ITM}, 2009, pp. 26--28.

\bibitem{Laurendeau2008}
C.~Laurendeau and M.~Barbeau, ``{Insider attack attribution using signal
  strength-based hyperbolic location estimation},'' \emph{Security and
  Communication Networks}, vol.~1, no. July, pp. 337--349, 2008.

\bibitem{Laurendeau2009}
------, ``{Probabilistic Localization and Tracking of Malicious Insiders Using
  Hyperbolic Position Bounding in Vehicular Networks},'' \emph{EURASIP Journal
  on Wireless Communications and Networking}, vol. 2009, no.~1, p. 128679,
  2009.

\bibitem{Strasser2008}
M.~Strasser, C.~P\"{o}pper, S.~Capkun, and M.~Cagalj, ``{Jamming-resistant Key
  Establishment using Uncoordinated Frequency Hopping},'' in \emph{2008 IEEE
  Symposium on Security and Privacy (sp 2008)}.\hskip 1em plus 0.5em minus
  0.4em\relax IEEE, May 2008, pp. 64--78.

\bibitem{Christina2009}
C.~P\"{o}pper, M.~Strasser, and S.~Capkun, ``{Jamming-resistant Broadcast
  Communication without Shared Keys},'' in \emph{Proceedings of the USENIX
  Security Symposium}, 2009, pp. 231--247.

\bibitem{Liu2010}
Y.~Liu, P.~Ning, H.~Dai, and A.~Liu, ``{Randomized Differential DSSS:
  Jamming-Resistant Wireless Broadcast Communication},'' \emph{2010 Proceedings
  IEEE INFOCOM}, pp. 1--9, Mar. 2010.

\bibitem{Finke2013a}
C.~Finke, J.~Butts, R.~Mills, and M.~Grimaila, ``{Enhancing the security of
  aircraft surveillance in the next generation air traffic control system},''
  \emph{International Journal of Critical Infrastructure Protection}, vol.~6,
  no.~1, pp. 3--11, Mar. 2013.

\bibitem{wessoncan}
K.~D. Wesson, T.~E. Humphreys, and B.~L. Evans, ``{Can Cryptography Secure Next
  Generation Air Traffic Surveillance?}'' Tech. Rep., Jan. 2014.

\bibitem{Samuelson2006}
K.~Samuelson, E.~Valovage, and D.~Hall, ``{Enhanced ADS-B Research},'' in
  \emph{2006 IEEE Aerospace Conference}.\hskip 1em plus 0.5em minus 0.4em\relax
  IEEE, 2006, pp. 1--7.

\bibitem{viggiano2010secure}
M.~Viggiano, E.~Valovage, K.~Samuelson, D.~Hall \emph{et~al.}, ``{Secure ADS-B
  authentication system and method},'' Jun 2010, {US Patent 7,730,307}.

\bibitem{schuchman2011automatic}
L.~Schuchman, ``{Automatic dependent surveillance system secure ADS-S},'' 2011.

\bibitem{Robinson2007}
R.~V. Robinson, M.~Li, S.~A. Lintelman, K.~Sampigethaya, R.~Poovendran, D.~V.
  Oheimb, and J.-U. Bu{\ss}er, ``{Impact of Public Key Enabled Applications on
  the Operation and Maintenance of Commercial Airplanes},'' in \emph{AIAA
  Aviation Technology Integration, and Operations (ATIO) Conference}, 2007.

\bibitem{Ziliang2010}
F.~Ziliang, P.~A.~N. Weijun, and W.~Yang, ``{A Data Authentication Solution Of
  ADS-B System Based On X.509 Certificate},'' in \emph{27th International
  Congress of the Aeronautical Sciences}, 2010.

\bibitem{Moody2000}
C.~Moody, ``{System Description for the Universal Access Transceiver},'' FAA,
  Tech. Rep. November, 2000.

\bibitem{Raya2005}
M.~Raya and J.-P. Hubaux, ``{The security of vehicular ad hoc networks},'' in
  \emph{Proceedings of the 3rd ACM workshop on Security of ad hoc and sensor
  networks - SASN '05}.\hskip 1em plus 0.5em minus 0.4em\relax New York, New
  York, USA: ACM Press, 2005, p.~11.

\bibitem{Raya2007}
------, ``{Securing vehicular ad hoc networks},'' \emph{Journal of Computer
  Security}, vol.~15, no.~1, pp. 39--68, 2007.

\bibitem{Parno}
B.~Parno and A.~Perrig, ``{Challenges in Securing Vehicular Networks},'' in
  \emph{Workshop on Hot Topics in Networks (HotNets-IV)}, 2005, pp. 1--6.

\bibitem{Zhang2010}
J.~Zhang and V.~Varadharajan, ``{Wireless sensor network key management survey
  and taxonomy},'' \emph{Journal of Network and Computer Applications},
  vol.~33, no.~2, pp. 63--75, Mar. 2010.

\bibitem{Perrig2000}
A.~Perrig, R.~Canetti, J.~D. Tygar, and D.~Song, ``{Efficient Authentication
  and Signing of Multicast Streams over Lossy Channels},'' in \emph{Security
  and Privacy, 2000. S\&P 2000. Proceedings. 2000 IEEE Symposium on}, 2000, pp.
  56--73.

\bibitem{Perrig2005}
A.~Perrig, D.~Song, and R.~Canetti, ``{The TESLA Broadcast Authentication
  Protocol},'' 2005.

\bibitem{Perrig2002}
A.~Perrig, R.~Szewczyk, J.~D. Tygar, V.~Wen, and D.~E. Culler, ``{SPINS:
  Security Protocols for Sensor Networks},'' \emph{Wireless networks 8},
  vol.~8, no.~5, pp. 521--534, 2002.

\bibitem{Liu}
D.~Liu, P.~Ning, S.~Zhu, and S.~Jajodia, ``{A Tree-Based µTESLA Broadcast
  Authentication for Sensor Networks}.''

\bibitem{Eldefrawy2010}
M.~H. Eldefrawy, M.~K. Khan, K.~Alghathbar, and E.-S. Cho, ``{Broadcast
  authentication for wireless sensor networks using nested hashing and the
  Chinese remainder theorem.}'' \emph{Sensors}, vol.~10, no.~9, pp. 8683--95,
  Jan. 2010.

\bibitem{Haas2009}
J.~J. Haas, Y.-C. Hu, and K.~P. Laberteaux, ``{Real-World VANET Security
  Protocol Performance},'' in \emph{GLOBECOM 2009 - 2009 IEEE Global
  Telecommunications Conference}.\hskip 1em plus 0.5em minus 0.4em\relax IEEE,
  Nov. 2009, pp. 1--7.

\bibitem{Hu2006}
Y.-C. Hu and K.~P. Laberteaux, ``{Strong VANET Security on a Budget},'' in
  \emph{Proceedings of Workshop on Embedded Security in Cars (ESCAR)}, vol.~06,
  2006, pp. 1--9.

\bibitem{Savvides2002}
A.~Savvides, H.~Park, and M.~B. Srivastava, ``{The bits and flops of the n-hop
  multilateration primitive for node localization problems},'' in
  \emph{Proceedings of the 1st ACM international workshop on Wireless sensor
  networks and applications - WSNA '02}.\hskip 1em plus 0.5em minus 0.4em\relax
  New York, New York, USA: ACM Press, 2002, p. 112.

\bibitem{Neven2005}
W.~Neven, T.~Quilter, R.~Weedon, and R.~Hogendoorn, ``{Wide Area
  Multilateration Report on EATMP TRS 131/04 Version 1.1},'' National Aerospace
  Laboratory NLR, Tech. Rep., 2005.

\bibitem{sayed2005network}
A.~H. Sayed, A.~Tarighat, and N.~Khajehnouri, ``Network-based wireless
  location: challenges faced in developing techniques for accurate wireless
  location information,'' \emph{Signal Processing Magazine, IEEE}, vol.~22,
  no.~4, pp. 24--40, 2005.

\bibitem{Xu2010}
N.~Xu, R.~Cassell, and C.~Evers, ``{Performance assessment of Multilateration
  Systems - A solution to nextgen surveillance},'' in \emph{Integrated
  Communications Navigation and Surveillance Conference (ICNS)}, 2010, pp.
  2--9.

\bibitem{HerreroJ.G.;BesadaPortasJ.A.;RodriguezF.J.Jimenez;Corredera1999}
J.~J. Herrero, J.~Portas, J.~C. R.~C. Corredera, J.~{Besada Portas}, and
  F.~Rodriguez, ``{ASDE and multilateration mode-S data fusion for location and
  identification on airport surface},'' in \emph{Radar Conference, 1999. The
  Record of the 1999 IEEE}, no.~34, 1999, pp. 315--320.

\bibitem{Johnson2012}
J.~Johnson, H.~Neufeldt, and J.~Beyer, ``{Wide area multilateration and ADS-B
  proves resilient in Afghanistan},'' in \emph{Integrated Communications,
  Navigation and Surveillance Conference (ICNS), 2012 IEEE}, 2012.

\bibitem{Kaune2012}
R.~Kaune, C.~Steffes, S.~Rau, W.~Konle, and J.~Pagel, ``{Wide area
  multilateration using ADS-B transponder signals},'' in \emph{Information
  Fusion (FUSION), 2012 15th International Conference on}, 2012, pp. 727--734.

\bibitem{Thomas2011a}
P.~Thomas, ``{North sea helicopter ADS-B/MLat pilot project findings},'' in
  \emph{Digital Communications-Enhanced Surveillance of Aircraft and Vehicles
  (TIWDC/ESAV), 2011 Tyrrhenian International Workshop on}, 2011, pp. 53--58.

\bibitem{Daskalakis2003}
A.~Daskalakis and P.~Martone, ``{A technical assessment of ADS-B and
  multilateration technology in the Gulf of Mexico},'' in \emph{Proceedings of
  the 2003 IEEE Radar Conference}.\hskip 1em plus 0.5em minus 0.4em\relax IEEE,
  2003, pp. 370--378.

\bibitem{Niles2012}
F.~A. Niles, R.~S. Conker, M.~B. El-Arini, D.~G. O'Laighlin, and D.~V. Baraban,
  ``{Wide Area Multilateration for Alternate Position, Navigation, and Timing
  (APNT)},'' MITRE-CAASD, Tech. Rep.

\bibitem{Galati2005}
G.~Galati, M.~Leonardi, P.~Magar\`{o}, and V.~Paciucci, ``{Wide area
  surveillance using SSR mode S multilateration: advantages and limitations},''
  in \emph{European Radar Conference (EURAD)}, 2005, pp. 225--229.

\bibitem{douceur2002sybil}
J.~R. Douceur, ``{The sybil attack},'' \emph{Peer-to-peer Systems}, pp.
  251--260, 2002.

\bibitem{Brands1994}
S.~Brands and D.~Chaum, ``{Distance-bounding protocols},'' in \emph{Advances in
  Cryptology - EUROCRYPT'93}, 1994, pp. 344--359.

\bibitem{Clulow2006}
J.~Clulow, G.~P. Hancke, M.~G. Kuhn, and T.~Moore, ``{So Near and Yet So Far:
  Distance-Bounding Attacks in Wireless Networks},'' in \emph{Security and
  Privacy in Ad-hoc and Sensor Networks}, 2006, pp. 83--97.

\bibitem{Cremers2012}
C.~Cremers, K.~B. Rasmussen, B.~Schmidt, and S.~Capkun, ``Distance hijacking
  attacks on distance bounding protocols,'' in \emph{Security and Privacy (SP),
  2012 IEEE Symposium on}.\hskip 1em plus 0.5em minus 0.4em\relax IEEE, 2012,
  pp. 113--127.

\bibitem{Song2008}
J.-H. Song, V.~W. Wong, and V.~C. Leung, ``Secure location verification for
  vehicular ad-hoc networks,'' in \emph{Global Telecommunications Conference,
  2008. IEEE GLOBECOM 2008. IEEE}.\hskip 1em plus 0.5em minus 0.4em\relax IEEE,
  2008, pp. 1--5.

\bibitem{Chiang2009}
J.~T. Chiang, J.~J. Haas, and Y.-C. Hu, ``{Secure and precise location
  verification using distance bounding and simultaneous multilateration},'' in
  \emph{Proceedings of the second ACM conference on Wireless network security -
  WiSec '09}.\hskip 1em plus 0.5em minus 0.4em\relax New York, New York, USA:
  ACM Press, 2009, p. 181.

\bibitem{Chiang2012}
J.~Chiang, J.~Haas, J.~Choi, and Y.~Hu, ``{Secure location verification using
  simultaneous multilateration},'' \emph{IEEE Transactions on Wireless
  Communications}, vol.~11, no.~2, pp. 584--591, 2012.

\bibitem{Ranganathan}
A.~Ranganathan, N.~O. Tippenhauer, B.~\v{S}kori\'{c}, D.~Singel, and
  S.~\v{C}apkun, ``{Design and Implementation of a Terrorist Fraud Resilient
  Distance Bounding System},'' in \emph{Computer Security--ESORICS 2012}, 2012,
  pp. 415--432.

\bibitem{Tippenhauer2009}
N.~O. Tippenhauer and S.~\v{C}apkun, ``{ID-Based Secure Distance Bounding and
  Localization},'' in \emph{Computer Security--ESORICS 2009}, 2009, pp.
  621--636.

\bibitem{Kalman1960a}
R.~E. Kalman, ``A new approach to linear filtering and prediction problems,''
  \emph{Journal of basic Engineering}, vol.~82, no.~1, pp. 35--45, 1960.

\bibitem{Welch1995}
G.~Bishop and G.~Welch, ``An introduction to the kalman filter,'' \emph{Proc of
  SIGGRAPH}, 2001.

\bibitem{Fox2003}
D.~Fox, J.~Hightower, L.~Liao, and D.~Schulz, ``{Bayesian Filters for Location
  Estimation},'' \emph{Pervasive Computing}, vol.~2, no.~3, 2003.

\bibitem{Krozel2004}
J.~Krozel, D.~Andrisani, M.~A. Ayoubi, T.~Hoshizaki, and C.~Schwalm,
  ``{Aircraft ADS-B Data Integrity Check},'' in \emph{AIAA 4th Aviation
  Technology, Integration and Operations (ATIO) Forum}, 2004, pp. 1--11.

\bibitem{Chan-tin2010}
E.~Chan-Tin, V.~Heorhiadi, N.~Hopper, and Y.~Kim, ``{The Frog-Boiling Attack:
  Limitations of Secure Network Coordinate Systems},'' \emph{ACM Transactions
  on Information and System Security (TISSEC)}, vol.~14, no.~3, p.~27, 2011.

\bibitem{Baud2006}
O.~Baud, N.~Honore, and O.~Taupin, ``{Radar / ADS-B data fusion architecture
  for experimentation purpose},'' in \emph{9th International Conference on
  Information Fusion}, 2006, pp. 1--6.

\bibitem{liu2013multi}
W.~Liu, J.~Wei, M.~Liang, Y.~Cao, and I.~Hwang, ``Multi-sensor fusion and fault
  detection using hybrid estimation for air traffic surveillance,''
  \emph{Aerospace and Electronic Systems, IEEE Transactions on}, vol.~49,
  no.~4, pp. 2323--2339, 2013.

\bibitem{Wei2003}
Y.-C. Wei, Y.-M. Chen, and H.-L. Shan, ``{Beacon-based Trust Management for
  Location Privacy Enhancement VANETs},'' in \emph{13th Asia-Pacific Network
  Operations and Management Symposium (APNOMS)}, 2011.

\bibitem{Yan2008}
G.~Yan, S.~Olariu, and M.~C. Weigle, ``{Providing VANET security through active
  position detection},'' \emph{Computer Communications}, vol.~31, no.~12, pp.
  2883--2897, Jul. 2008.

\bibitem{smith2008method}
A.~E. Smith, ``{Method and apparatus for improving ADS-B security},'' Sep 2008,
  {US Patent 7,423,590}.

\bibitem{Xiao}
B.~Xiao, B.~Yu, and C.~Gao, ``{Detection and Localization of Sybil Nodes in
  VANETs},'' in \emph{International Conference on Mobile Computing and
  Networking: Proceedings of the 2006 workshop on Dependability issues in
  wireless ad hoc networks and sensor networks}, 2006, pp. 1--8.

\bibitem{Leinmuller2006}
T.~Leinmuller, E.~Schoch, and F.~Kargl, ``{Position verification approaches for
  vehicular ad-hoc networks},'' \emph{IEEE Wireless Communications}, vol.~13,
  no.~5, pp. 16--21, Oct. 2006.

\bibitem{Martinovic2009}
I.~Martinovic, P.~Pichota, and J.~B. Schmitt, ``{Jamming for Good: A Fresh
  Approach to Authentic Communication in WSNs},'' in \emph{WiSec '09:
  Proceedings of the second ACM conference on Wireless network security}, 2009,
  pp. 161--168.

\bibitem{Wilhelm2011}
M.~Wilhelm and I.~Martinovic, ``{WiFire: A Firewall for Wireless Networks},''
  \emph{ACM SIGCOMM Computer Communication Review}, vol.~4, no.~4, pp.
  456--457, 2011.

\bibitem{Schaefer14}
M.~Sch{\"a}fer, M.~Strohmeier, V.~Lenders, I.~Martinovic, and M.~Wilhelm,
  ``{Bringing Up OpenSky: A Large-scale ADS-B Sensor Network for Research},''
  in \emph{ACM/IEEE International Conference on Information Processing in
  Sensor Networks (IPSN)}, Apr. 2014.

\end{thebibliography}

\end{document}